\documentclass[journal]{IEEEtran}
\usepackage{amsmath,amsfonts}
\usepackage{comment}
\usepackage{multirow}
\usepackage{graphicx}
\usepackage{bbold}
\ifCLASSINFOpdf
\else
\fi
\newcommand{\T}{\mathcal{T}}
\newcommand{\K}{\mathcal{K}}

\renewcommand{\vec}[1]{\mathbf{#1}}
\newcommand{\mat}[1]{\boldsymbol{\mathsf{#1}}} 

\newcommand\notsotiny{\@setfontsize\notsotiny\@vipt\@viipt}

\begin{document}
%
\title{Quasi-Helmholtz Calder\'on Multiplicative Preconditioning for Higher-Order Global Multi-Trace Integral Equations}
%
%
%

\author{Cedric~M\"{u}nger,~\IEEEmembership{Member,~IEEE},  Alessandro Zuccotti,~\IEEEmembership{Student Member,~IEEE},  Van Chien Le, ~\IEEEmembership{Member,~IEEE}, and~Kristof~Cools,~\IEEEmembership{Member,~IEEE} 
\thanks{This work was supported by the European Research Council (ERC) under the European Union's Horizon 2020 Research and Innovation Program under Grant 101001847 and the Special Research Fund (BOF) of Ghent University under Grant BOF.PDO.2024.0016.01.}
\thanks{C. M\"{u}nger, A. Zuccotti, V.~C. Le, and K. Cools are with IDLab, Department of Information Technology at Ghent University - imec, Ghent, Belgium (Corresponding author: Cedric M\"{u}nger e-mail: cedric.muenger@ugent.be).}}

%
%

\markboth{IEEE Transactions on Antennas and Propagation}%
{Shell \MakeLowercase{\textit{et al.}}: Bare Demo of IEEEtran.cls for IEEE Journals}
%



\maketitle

\begin{abstract}
The paper presents a higher-order global multi-trace integral equation for time-harmonic electromagnetic scattering by composite objects. The higher-order multi-trace formulation is preconditioned with a Calder\'{o}n multiplicative preconditioner using higher-order quasi-Helmholtz projectors. The higher-order quasi-Helmholtz projectors separate the solenoidal and non-solenoidal components of the basis functions. Separate access to the Helmholtz components circumvents the explicit inversion of an ill-conditioned mixed Gram matrix. This enables the application of Calder\'{o}n multiplicative preconditioning without refining the mesh and resorting to dual basis functions. Furthermore, it enables low-frequency stabilization, as the solenoidal and non-solenoidal components can be rescaled individually. The higher-order quasi-Helmholtz projectors are computed iteratively, enabling iterative solvers to efficiently solve the preconditioned matrix system, yielding accurate solutions in a few dozen iterations. Numerical experiments are conducted for composite dielectric bodies, confirming the effectiveness of the proposed preconditioner for the global multi-trace formulation discretized with higher-order basis functions for dense meshes and at very low frequencies. 
\end{abstract}


%
\IEEEpeerreviewmaketitle

\section{Introduction}
%
%
%
%
\IEEEPARstart{M}{ulti-Trace} integral equations are a powerful tool for modeling scattering of time-harmonic electromagnetic waves on composite objects, as the multi-trace technique readily allows the application of Calder\'on multiplicative preconditioning to composite structures with junctions \cite{chu_surface_2003, claeys_electromagnetic_2012, lasisi_fast_2022}. However, if higher-order basis functions, such as the Graglia--Wilton--Peterson (GWP) basis functions \cite{graglia_higher_1997}, are used in the discretization, Calder\'on preconditioning based on explicit dual basis functions is more challenging to construct, as it will require higher-order dual functions, akin to the Buffa--Christiansen (BC) functions \cite{buffa_dual_2007} that are used in conjunction with Rao--Wilton--Glisson (RWG) \cite{rao_electromagnetic_1982} in the zeroth-order case \cite{andriulli_multiplicative_2008}.

There have been some efforts to generalize the BC dual functions beyond the lowest order. For example, in \cite{valdes_high-order_2011}, a barycentric refinement approach using higher-order basis functions is used to numerically construct higher-order dual basis functions. 

Alternative Calder\'{o}n multiplicative preconditioning strategies exist \cite{christiansen_preconditioner_2003, adrian_refinement-free_2019}, which do not rely on mesh refinement and dual basis functions. These alternative approaches leverage the Helmholtz decomposition in one form or another, separating the surface current into solenoidal and nonsolenoidal components. 

Recently, a similar multiplicative Calder\'on preconditioning scheme, also based on the Helmholtz decomposition, has been developed for the higher-order EFIE \cite{bourhis_novel_2024}. It leverages higher-order quasi-Helmholtz projectors, making this a relatively easy-to-use Calder\'on preconditioning technique for basis functions of arbitrary order.

Higher-order quasi-Helmholtz projectors \cite{merlini_unified_2019} make it possible to separate a basis of any order into solenoidal and non-solenoidal functions. This enables the separate mapping of the solenoidal or non-solenoidal functions of order $p$ to non-solenoidal or solenoidal functions of order higher than $p$ \cite{bourhis_high-order_2024}. Thus, the discrete duality between spaces of different orders can be achieved without explicitly inverting an ill-conditioned mixed Gram matrix, thereby enabling the application of Calder\'on multiplicative preconditioning without resorting to mesh refinement and dual basis functions. 

This contribution utilizes higher-order quasi-Helmholtz projectors to construct block-diagonal Calder\'on preconditioners for higher-order discretizations of global multi-trace formulations. It has been shown that retaining only the preconditioner's diagonal contributions lowers the construction cost of the preconditioner while still obtaining the desired preconditioning effect \cite{lasisi_fast_2022}. Furthermore, this work introduces a new alternative method for efficiently computing the action of higher-order quasi-Helmholtz projectors. The alternative approach does not require a specialized solver or explicit computation of a pseudo-inverse. Instead, it is built with standard tools that require no to little additional work to deploy. At its core, it uses a well-studied preconditioning technique for mixed finite element problems \cite{powell_optimal_2003}, which is sufficient for this approach to be combined with a standard iterative solver such as GMRES \cite{saad_gmres_1986}.  


The paper is structured as follows: Section \ref{sec:mt} and \ref{sec:hoqhp} introduce the relevant concepts and notations for global multi-trace formulations and higher-order quasi-Helmholtz projectors, respectively. In Section \ref{sec:cmp}, those techniques are combined to formulate a higher-order multi-trace integral equation amenable to block-diagonal Calder\'on multiplicative preconditioning, yielding a well-conditioned linear system. In section \ref{sec:impl}, essential implementation details are covered to enable an iterative solver to solve the linear system efficiently. In particular, it will be demonstrated how the action of the higher-order quasi-Helmholtz projectors can be computed efficiently (without involving an expensive pseudo-inverse). In section \ref{sec:lfstab}, low-frequency stabilization is discussed, and it is shown that the modified scheme is stable at very low frequencies. Last but not least, section \ref{sec:num} presents a series of numerical experiments that corroborate the well-conditioning of the higher-order multi-trace formulation with the quasi-Helmholtz Calder\'on multiplicative preconditioning. Finally, the last section includes some concluding remarks and perspectives.

\section{Global Multi-Trace Formulation}
\label{sec:mt}
For clarity and space constraints, we restrict the following discussion to two domains, but the global multi-trace formulation can be readily extended to any number of touching domains.

\begin{figure}[ht]
  
    \centering
    \includegraphics[width=0.75\linewidth, height=4cm]{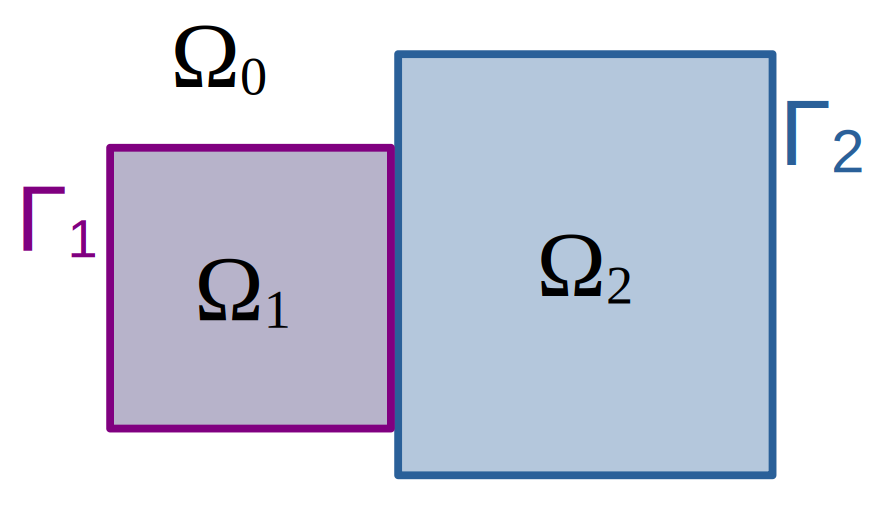}
    \caption{Touching domains $\Omega_1$ and $\Omega_2$ embedded in $\Omega_0$ together with their respective boundaries $\Gamma_1$ and $\Gamma_2$}
      \label{fig:domains}
\end{figure}

Consider two domains $\Omega_1 \subset \mathbb{R}^3$ and  $\Omega_2 \subset \mathbb{R}^3$. $\Gamma_1:= \partial\Omega_1$ and $\Gamma_2:= \partial\Omega_2$ are the boundaries of $\Omega_1$ and $\Omega_2$, respectively. The two domains do not overlap but may touch each other, i.e., $\Omega_1 \cap \Omega_2 = \emptyset $ but possibly $\Gamma_1 \cap \Gamma_2 \neq \emptyset $. The background is denoted as $\Omega_0:= \mathbb{R}^3 \setminus \overline{\Omega_1 \cup \Omega_2}$, see Fig \ref{fig:domains}. The impedance and the wave number of the material filling $\Omega, \Omega_1,$ and $\Omega_2$  are denoted as $\eta_0, \eta_1,$ and $ \eta_2$,  $\kappa_0, \kappa_1,$ and $\kappa_2 $, respectively.

The Maxwell single and double layer boundary integral operators are defined as follows:
\begin{equation}
    \small
    \label{eq:slp}
    \begin{split}
            \mathcal{T}_{lm}^{(s)}[\vec{j}](\vec{r}_l) = &-i\kappa_s \, \vec{n}_l \times \int_{\Gamma_m} \frac{e^{-i\kappa_s R}}{4\pi R} \vec{j}(\vec{r}_m) d\vec{r}_m \\&+ \frac{1}{i\kappa_s} \vec{n}_l \times \nabla_l \int_{\Gamma_m} \frac{e^{-i\kappa_s R}}{4\pi R} \nabla_m \cdot \vec{j}(\vec{r}_m) d\vec{r}_m \quad \vec{r}_l \in \Gamma_l,
    \end{split}
\end{equation}
\begin{equation}
    \small
    \label{eq:dlp}
    \mathcal{K}_{lm}^{(s)}[\vec{j}](\vec{r}_l) = \vec{n}_l \times \text{p.v.} \int_{\Gamma_m} \nabla \, \frac{e^{-i\kappa_s R}}{4\pi R} \times \vec{j}(\vec{r}_m) d\vec{r}_m \quad \vec{r}_l \in \Gamma_l,
\end{equation}
where $R = \vert \vec{r}_l-\vec{r}_m\vert$, $i = \sqrt{-1}$ is the imaginary unit, p.v. stands for the Cauchy principal value, $l$ and $m$ denote the indices of the test and trial spaces, $\vec{n}_l$ the outward pointing normal vector on surface $\Gamma_l$, and $(s)$ specifies the material parameter.

The global multi-trace formulation is an extension of the PMCHWT \cite{poggio_chapter_1973} to multiple domains. For the two dielectric domains $\Omega_1$ and $\Omega_2$, the global multi-trace formulation is shown in \eqref{eq:mt}, where \eqref{eq:slp} and \eqref{eq:dlp} are used in combination with the electic and magnetic surface currents $\vec{j}_1(\vec{r})$, $\vec{m}_1(\vec{r})$, $\vec{j}_2(\vec{r})$, and $\vec{m}_2(\vec{r})$ on $\Gamma_1$ and $\Gamma_2$, and the incident electric and magnetic fields  $\vec{E}^{inc}(\vec{r})$ and $\vec{H}^{inc}(\vec{r})$.

\begin{figure*}

\begin{align} \label{eq:mt}
\begin{pmatrix}
 \eta_0 \T^{(0)}_{11} +  \eta_1 \T^{(1)}_{11}   & -\K^{(0)}_{11}-\K^{(1)}_{11}     & \eta_0 \T^{(0)}_{12}        & -\K^{(0)}_{12}+\frac{1}{2}   \\
 \K^{(0)}_{11}+\K^{(1)}_{11} &  \frac{1}{\eta_0} \T^{(0)}_{11} + \frac{1}{\eta_1} \T^{(1)}_{11}  & \K^{(0)}_{12}-\frac{1}{2} & \frac{1}{\eta_0} \T^{(0)}_{12} \\
 \eta_0 \T^{(0)}_{21}        & -\K^{(0)}_{21}+\frac{1}{2}     & \eta_0 \T^{(0)}_{22} + \eta_2 \T^{(2)}_{22}       & -\K^{(0)}_{22}-\K^{(2)}_{22}   \\
 \K^{(0)}_{21}-\frac{1}{2} &  \frac{1}{\eta_0} \T^{(0)}_{21}  & \K^{(0)}_{22}+\K^{(2)}_{22} & \frac{1}{\eta_0} \T^{(0)}_{22}+\frac{1}{\eta_2} \T^{(2)}_{22} \\
\end{pmatrix}
 \begin{pmatrix}
 \vec{j}_1(\vec{r}) \\ \vec{m}_1(\vec{r}) \\  \vec{j}_2(\vec{r}) \\ \vec{m}_2(\vec{r}) \\
\end{pmatrix}
=  
 - \begin{pmatrix} \vec{E}^{inc}(\vec{r}) \times \vec{n_1} \\ \vec{n_1} \times \vec{H}^{inc}(\vec{r}) \\ \vec{E}^{inc}(\vec{r}) \times \vec{n_2} \\ \vec{n_2} \times \vec{H}^{inc}(\vec{r}) 
\end{pmatrix}
\end{align}

\end{figure*}

More details on global multi-trace formulations can be found in \cite{chu_surface_2003, claeys_electromagnetic_2012, lasisi_fast_2022}.

\section{Discrete higher-Order Quasi-Helmholtz Projectors}
\label{sec:hoqhp}

Let $\mathcal{GWP}_{p} = \{\mathcal{GWP}_{p,1},\cdots,\mathcal{GWP}_{p,N}\}$ be the set of div-conforming Graglia--Wilton--Peterson (GWP) basis functions of order $p \geq 0$ \cite{graglia_higher_1997}. $\mathcal{P}_{p}$ and $\mathcal{L}_{q}$ are the sets of scalar discontinuous and continuous Lagrange basis functions of order $p \geq 0$ and $q \geq 1$, respectively. The star matrix $\Sigma_{p}$ of order $p$ has the following entries \cite{merlini_unified_2019}:
\begin{equation}
    \Sigma_{p,mn} = \langle \operatorname{div } \mathcal{GWP}_{p,m}, \mathcal{P}_{p,n} \rangle,
\end{equation}
where the angled brackets denote the $L^2$ inner product.  This allows for defining the order $p$ quasi-Helmholtz projector
\begin{equation}
    \label{eq:psigma}
    \mathrm{P}^\Sigma_p = \Sigma_p ( \Sigma_p^T G_p^{-1} \Sigma_p)^+ \Sigma_p^T,    
\end{equation}
where $^+$ denotes the pseudo-inverse, and $G_p^{-1}$ is the inverse of GWP Gram matrix ($G_{p,mn}= \langle\mathcal{GWP}_{p,m},\mathcal{GWP}_{p, n} \rangle$). A complementary projector to $\mathrm{P}^\Sigma_p$ can be defined as $\mathrm P^{\Lambda H}_p = G_p - \mathrm{P}^\Sigma_p$. By construction, it holds that $\mathrm P^{\Lambda H}_p G_p^{-1} \mathrm{P}^\Sigma_p = 0$. Furthermore, they have the following properties \cite{bourhis_novel_2024}:
\begin{equation}
    \mathrm P^{\Lambda H}_p  G_p^{-1} T_{p\vert h} = T_{p\vert h}  G_p^{-1}\mathrm P^{\Lambda H}_p = 0,
\end{equation}
where $T_{p\vert h}$ is the hypersingular part of the Maxwell single layer operator, discretized using order $p$ test and basis functions.
In addition, we have
\begin{equation}
    \mathrm P^{\Lambda H}_p  G_p^{-1} K_{p\vert 0}  G_p^{-1} \mathrm P^{\Lambda H}_p = 0,
\end{equation}
where $ K_{p\vert 0} $ is the static contribution ($\kappa \to 0$) of the double layer operator \cite{chen_analysis_2001}. 

There exists another pair of projectors based on the generalized loop matrix
\begin{equation}
    \Lambda_{p,mn} = \langle \mathcal{GWP}_{p,m}, \operatorname{\bf curl} \mathcal{L}_{p+1,n} \rangle.
\end{equation}
The quasi-Helmholtz projectors based on $\Lambda_p$ are defined as
\begin{equation}
    \label{eq:plambda}
    \mathbb P^\Lambda_p = \Lambda_p ( \Lambda_p^T G_p^{-1} \Lambda_p)^+ \Lambda_p^T,    
\end{equation}
and 
\begin{equation}
    \mathbb P^{\Sigma H}_p =  G_p - \mathbb \mathrm P^\Lambda_p. 
\end{equation}
This second pair of quasi-Helmholtz projectors will be essential later when a low-frequency stable formulation for multiply connected geometries is developed.

\section{Quasi-Helmholtz Calder\'on Multiplicative Preconditioning}
\label{sec:cmp}

For Calder\'on multiplicative preconditioning (CMP) with higher-order basis functions, a dual space to the higher-order primal space is needed. There are different ways to construct such a space for a higher-order basis. A higher-order dual space can be explicitly constructed from the barycentric refinement of the mesh \cite{valdes_high-order_2011}. Or a space of order higher than the primal space can be used \cite{christiansen_preconditioner_2003}. In this work, we follow the latter approach, with a space of order $r=p+2$, which was found to be optimal for the EFIE \cite{bourhis_high-order_2024}.  
  
For CMP to work, it is crucial that the dimension of the primal solenoidal subspace at least matches that of the dual nonsolenoidal subspace and that the dimension of the primal nonsolenoidal subspace matches that of the dual solenoidal subspace  \cite{valdes_high-order_2011}. 

With the help of higher-order quasi-Helmholtz projectors, it is possible to construct a discrete higher-order duality pairing that maps the solenoidal or nonsolenoidal functions of GWP space of order $p$ on the nonsolenoidal or solenoidal functions of GWP space of higher order $r>p$, which then are able to act as the dual function to the GWP functions of order $p$. During this projection, the solenoidal and nonsolenoidal components are rotated separately and projected to the nonsolenoidal and solenoidal components of the GWP functions of order $r$, respectively. In discretized form, these operations can be written as follows \cite{bourhis_novel_2024}:
\begin{align}
    \Theta^\Sigma_{rp} =  \mathrm P^\Sigma_r G_r^{-1} N_{rp} G_p^{-1}\mathrm P^{\Lambda H}_p, \\
    \Theta^{\Lambda H}_{rp} = \mathrm P^{\Lambda H}_r  G_r^{-1} N_{rp}  G_p^{-1} \mathrm{P}^\Sigma_p.
\end{align}
Here, $N_{rp}$ is a mixed Gram matrix with entries $N_{rp,mn} = \langle \mathcal{GWP}_{r,m},\vec{n} \times\mathcal{GWP}_{p,n} \rangle$. $\Theta^\Sigma_{rp} $ maps solenoidal functions of order $p$ on nonsolenoidal functions of order $r$ and  $\Theta^{\Lambda H}_{rp}$ maps nonsolenoidal functions of order $p$ on solenoidal functions of order $r$.

$\Theta^\Sigma_{rp}$ and $\Theta^{\Lambda H}_{rp}$ can be used  to build the preconditioner $\mathbb{T}_r$ \cite{bourhis_novel_2024}:
 \begin{equation}
       \mathbb{T}^{(s)}_r =  G_p^{-1}(\Theta^\Sigma_{pr} + \Theta^{\Lambda H}_{pr}) G_r^{-1} T^{(s)}_r  G_r^{-1}(\Theta^\Sigma_{rp} + \Theta^{\Lambda H}_{rp})  G_p^{-1}.
 \end{equation}

The discretized, block-preconditioned version of the higher-order multi-trace system \eqref{eq:mt} is given in \eqref{eq:precond_mt}. Due to its block structure, it can be easily extended to an arbitrary number of domains. $T^{(s)}_P$ and $K^{(s)}_p$ are the Maxwell single and double layer operators discretized using order-$p$ rotated ($\vec{n}\times$) GWPs as test functions and  order-$p$ GWPs as trial functions, and $(s)$ denotes the material properties. For readability, the boundary indices are omitted, as they can be deduced from the positions within the system.  $N_p$ is the mixed Gram matrix with entries $N_{p,mn} = \langle\mathcal{GWP}_{p,m},  \vec{n} \times\mathcal{GWP}_{p,n} \rangle$. $\vec{j}_1$, $\vec{m}_1$, $\vec{j}_2$, and $\vec{m}_2$ are the coefficient vectors. The discretized right hand side vectors  $\vec{e}_p^{inc}$  with entries $\vec{e}_{p,n}^{inc} = \langle \vec{n} \times \mathcal{GWP}_{p,n},   \vec{E}^{inc} \times \vec{n} \rangle$ and $\vec{h}_p^{inc}$  with entries $\vec{h}_{p,n}^{inc} = \langle   \vec{n} \times \mathcal{GWP}_{p,n} ,\vec{n} \times \vec{H}^{inc}\rangle$.

\begin{figure*}

 \begin{equation}
  \label{eq:precond_mt}
    \fontsize{6pt}{6pt}\selectfont
     \begin{bmatrix}
         \mathbb{T}^{(0)}_r \\
         & \mathbb{T}^{(0)}_r \\
          && \mathbb{T}^{(0)}_r \\
          &&&   \mathbb{T}^{(0)}_r
     \end{bmatrix}
     \begin{bmatrix}
         \eta_0 T^{(0)}_p +  \eta_1 T_p^{(1)}   & -K^{(0)}_p-K_p^{(1)}  & \eta_0 T^{(0)}_p& -K^{(0)}_p-\frac{1}{2}N_p\\
        K^{(0)}_p+K_p^{(1)}  & \frac{1}{\eta_0} T^{(0)}_p +  \frac{1}{\eta_1} T_p^{(1)} & K^{(0)}_p+\frac{1}{2}N_p& \frac{1}{\eta_0}T^{(0)}_p\\
          \eta_0 T^{(0)}_p& -K^{(0)}_p-\frac{1}{2}N_p&\eta_0 T^{(0)}_p +  \eta_2 T_p^{(2)}   & -K^{(0)}_p-K_p^{(2)}  \\
        K^{(0)}_p+\frac{1}{2}N_p& \frac{1}{\eta_0}T^{(0)}_p&K^{(0)}_p+K_p^{(2)}  & \frac{1}{\eta_0} T^{(0)}_p +  \frac{1}{\eta_2} T_p^{(2)}
     \end{bmatrix}
     \begin{bmatrix}
         \vec{j}_1 \\
         \vec{m}_1 \\
         \vec{j}_2 \\
         \vec{m}_2
     \end{bmatrix}
     = 
   -\begin{bmatrix}
         \mathbb{T}^{(0)}_r \\
         & \mathbb{T}^{(0)}_r \\
          && \mathbb{T}^{(0)}_r \\
          &&&   \mathbb{T}^{(0)}_r
     \end{bmatrix}
     \begin{bmatrix}
         \vec{e}_p^{inc} \\
         \vec{h}_p^{inc} \\
         \vec{e}_p^{inc} \\
         \vec{h}_p^{inc}
     \end{bmatrix}
 \end{equation}

 \end{figure*}

\section{Implementation Details}
\label{sec:impl}
The computation of the higher-order projectors \eqref{eq:psigma} and \eqref{eq:plambda} requires the numerical (pseudo-)inversion of the products of the generalized star and loop matrices.  
In the zeroth-order case, dedicated algorithms like algebraic multigrid \cite{livne_lean_2012, napov_algebraic_2012} can be used to compute the pseudo-inverse iteratively \cite{andriulli_loop-star_2012}. Using standard algebraic multigrid in a black-box fashion for the higher-order projector as defined in \eqref{eq:psigma} is no longer possible because the matrix that needs to be pseudo-inverted is no longer sparse due to the inverse Gram matrix between the higher-order star or loop matrices. There are alternative definitions of higher-order quasi-Helmholtz projectors \cite{bourhis_high-order_2024}, for which higher-order algebraic multigrid methods \cite{napov_algebraic_2014} still work. However, such specialized solvers are not widely available in standard numerical tools for use as black-box solvers.

In this work, an alternative approach is used to compute the higher-order quasi-Helmholtz projectors without resorting to algebraic multigrid. Instead, the quasi-Helmholtz projectors are recast into an equivalent saddle-point problem, which can be solved efficiently (with an appropriate preconditioner) with a standard iterative solver such as GMRES \cite{saad_gmres_1986}. 

\subsection{Iterative Computation of Projectors}
\label{sec:qh}
 We observe that an explicit expression for the quasi-Helmholtz projectors is unnecessary when an iterative solver is used to solve the problem involving the projectors. Only the projector's action on a coefficient vector is needed if the entire system is never explicitly inverted. Thus, instead of explicitly computing the projector matrices, which would require the pseudo-inversion of $\Sigma_p^T G_p^{-1} \Sigma_p$, \eqref{eq:psigma} can be rewritten involving a saddle-point problem

\begin{equation}
\label{eq:saddlepointA}
     \mathrm{P}^\Sigma_p \vec{j} 
    =
    \begin{bmatrix}
        0 & \Sigma_p
    \end{bmatrix}
    \begin{bmatrix}
        G_p &\Sigma_p \\
        \Sigma_p^T &  0 
    \end{bmatrix}^{-1}
    \begin{bmatrix}
        G_p \\
        0
    \end{bmatrix}
    \vec{j}
\end{equation}

If the inverse matrix in \eqref{eq:saddlepointA} is computed through the Schur complement, the equivalence between evaluating the right-hand side in \eqref{eq:saddlepointA} and applying the projector \eqref{eq:psigma} to $\vec{j}$ can be easily verified. 

The same saddle point problem can also be used to compute the action of the complementary projector $\mathrm P^\Lambda_p$

\begin{equation}
\label{eq:saddlepointB}
     \mathrm P^{\Lambda H}_p \vec{j} 
    =
    \begin{bmatrix}
        G_p & 0
    \end{bmatrix}
    \begin{bmatrix}
        G_p &\Sigma_p \\
        \Sigma_p^T &  0 
    \end{bmatrix}^{-1}
    \begin{bmatrix}
        G_p \\
        0
    \end{bmatrix}
    \vec{j}.
\end{equation}

This saddle-point problem also arises in the preconditioning scheme of \cite{christiansen_preconditioner_2003}.


\subsection{Solving the Saddle-Point Problem}
To evaluate \eqref{eq:saddlepointA} or \eqref{eq:saddlepointB}, the inverse of the saddle-point matrix has to be computed. 

The same kind of saddle-point system also appears in other fields; for instance, the mixed discretization of the Poisson problem \cite{raviart_mixed_1977} or the finite element discretization of the Darcy equation results in a saddle-point problem \cite{masud_stabilized_2002}.
Thus, saddle-point problems with the same structure as the one appearing in \eqref{eq:saddlepointA} and \eqref{eq:saddlepointB} have already been extensively studied, and efficient preconditioners exist that enable the inversion of the saddle-point matrix using an iterative solver in only a few iterations, rather than computing the expensive pseudo-inverse. 

This work uses a block-diagonal preconditioner proposed in \cite{powell_optimal_2003} when evaluating the saddle-point problem \eqref{eq:saddlepointA} iteratively. The saddle point matrix 
\begin{equation}
    \begin{bmatrix}
        G_p & \Sigma_p \\
        \Sigma_p^T & 0
    \end{bmatrix}
\end{equation}
can be preconditioned with the inverse of the following block diagonal matrix:

\begin{equation}
\label{eq:precondA}
 P_{div} =\begin{bmatrix}
     G_p + G^{div}_p & 0 \\
     0 & P_p 
 \end{bmatrix},    
\end{equation}
where $G^{div}_{p,mn} = \langle \text{div }\mathcal{GWP}_{p,m}, \text{div } \mathcal{GWP}_{p,n} \rangle$ and  $P_p$ is the Gram matrix of the discontinuous Largange basis functions of order $p$ ($P_{p,mn}= \langle \mathcal{P}_{p,m}, \mathcal{P}_{p,n} \rangle$).

Besides the projectors \eqref{eq:saddlepointA} and \eqref{eq:saddlepointB} based on the star matrix $\Sigma$, projectors can also be defined based on the loop matrix $\Lambda$. They can be computed iteratively in the same manner as their $\Sigma$-based counterparts, except that the arising saddle-point
problem 

\begin{equation}
\label{eq:saddlepointloopA}
     \mathbb{P}^\Lambda_p \vec{j} 
    =
    \begin{bmatrix}
        0 & \Lambda_p
    \end{bmatrix}
    \begin{bmatrix}
        G_p &\Lambda_p \\
        \Lambda_p^T &  0 
    \end{bmatrix}^{-1}
    \begin{bmatrix}
        G_p \\
        0
    \end{bmatrix}
    \vec{j}
\end{equation}

requires another preconditioner. The inverse of the following matrix has been found to work well for the saddlepoint matrix in \eqref{eq:saddlepointloopA}:

\begin{equation}
\label{eq:precondB}
    P_{curl} = \begin{bmatrix}
     G_p  & 0 \\
     0 & L_l + L^{curl}_l 
      \end{bmatrix},
\end{equation}
where $L_l$ is the Gram matrix of the continuous Lagrange basis functions of order $l=p+1$ ($L_{l,mn}= \langle \mathcal{L}_{l,m},\mathcal{L}_{l,n} \rangle$)  and $L^{curl}_{l,mn} = \langle \operatorname{\bf curl} \mathcal{L}_{l,m}, \operatorname{\bf curl }\mathcal{L}_{l,n} \rangle$.

Since all the matrices involved in $P_{div}$ and $P_{curl}$ are sparse and symmetric, a Cholesky decomposition is used to compute the blockwise inverse of the preconditioner directly. This Cholesky decomposition is performed only once at the beginning of the solution process; in the actual iteration of solving \eqref{eq:saddlepointA} or \eqref{eq:saddlepointloopA}, only the backward and forward substitution on the factorization is carried out.

\paragraph*{Note} The sparse matrices in \eqref{eq:precondA} and \eqref{eq:precondB} are defined per domain. The Cholesky decomposition is thus performed on a problem that is significantly smaller than the overall problem, assuming that all domains yield a similar number of unknowns.  If one or more domains are much larger than the others, it might be more economical to replace the explicit Cholesky decomposition with another approach to reduce computational cost.


\section{Low Frequency Stabilazation}
\label{sec:lfstab}
It is well known that integral equations such as the EFIE, MFIE, and PMCHWT suffer from low-frequency breakdown (sometimes also referred to as low-frequency ill-conditioning) \cite{chen_analysis_2001,zhao_integral_2000}. Quasi-Helmholtz projectors have been shown to effectively treat low-frequency breakdown, even in multiply connected geometries \cite{merlini_magnetic_2020}. In combination with CMP, they yield a method that is stable and well-conditioned at very low frequencies.  
Since the analysis of the low-frequency breakdown has already been carried out in detail for EFIE, MFIE, and PMCHWT in \cite{andriulli_loop-star_2012,bogaert_low-frequency_2014,beghein_low-frequency_2017} before, we restrict the discussion here to what is specific to this context. 

In the Helmholtz decomposition, the basis is decomposed into three parts: the solenoidal local loops $\vec{l}$, stars $\vec{s}$, and the harmonic functions (global loops) $\vec{g}$

\begin{equation}
    \vec{j} = \Lambda \vec{l} + H \vec{g} + \Sigma \vec{s}.
\end{equation}
Thus, the loop-star transformation matrix can be defined as:
\begin{equation}
    A = (\Lambda, H, \Sigma).
\end{equation}
where the matrices $\Lambda$, $\Sigma$, and $H$ are the loop, star, and discrete harmonic function matrices, respectively.

The following low-frequency scaling holds for any single and double layer operator $T$ and $K$ \cite{andriulli_loop-star_2012,bogaert_low_2011}:

\begin{equation}
\label{eq:lowT}
 A^T T A = \mathcal{O} \begin{pmatrix}
     \kappa & \kappa & \kappa \\
     \kappa & \kappa & \kappa \\
     \kappa & \kappa & \kappa^{-1}
 \end{pmatrix},    
\end{equation}

\begin{equation}
\label{eq:lowK}
 A^T K A = \mathcal{O} \begin{pmatrix}
     \kappa^2 & \kappa^2 & 1 \\
     \kappa^2 & 1 & 1 \\
     1 & 1 &  1
 \end{pmatrix}.  
\end{equation}

For the right-hand side excitation in the form of a plane wave, the scaling is \cite{beghein_low-frequency_2017}:

\begin{equation}
 A^T \vec{e}^{inc}= \mathcal{O} \begin{pmatrix}
     \kappa & \kappa & 1
 \end{pmatrix}.    
\end{equation}

With \eqref{eq:lowT} and \eqref{eq:lowK}, the low-frequency analysis of the multi-trace formulation \eqref{eq:mt} can be carried out. Since the diagonal and off-diagonal blocks of the global multi-trace formulation have identical scaling, we show only one $2\times2$ block of the system, which has a scaling similar to that of the PMCHWT. The algebraic structure of off-diagonal entries in the $2\times2$ off-diagonal blocks contributions can be deduced from the low-frequency analysis of the MFIE \cite{bogaert_low-frequency_2014}.

The low-frequency scaling of the diagonal and off-diagonal $2\times2$ blocks in \eqref{eq:precond_mt} is the following:

\begin{equation}
    \mathcal{O}\begin{pmatrix}
     1 & 1 & 1 & \kappa & \kappa^{-1} & \kappa^{-1} \\
     1 & 1 & 1 & \kappa & \kappa^{-1} & \kappa^{-1} \\
     \kappa^2 & \kappa^2 & 1 & \kappa & \kappa & \kappa \\
     \kappa & \kappa^{-1} & \kappa^{-1} & 1 & 1 & 1 \\
     \kappa & \kappa^{-1} & \kappa^{-1} & 1 & 1 & 1\\
     \kappa & \kappa & \kappa & \kappa^2 & \kappa^2 & 1
    \end{pmatrix}
\end{equation}
From this, it is clear that a breakdown will occur at very low frequencies as $\kappa$ approaches $0$. 

The low-frequency breakdown can be addressed by rescaling the coefficients and the block preconditioner $\mathbb{T}_r$ in \eqref{eq:precond_mt}. Here, the preconditioned system is low freqeuncy stabilized directly, compared to \cite{beghein_low-frequency_2017} where the preconditioner and the system are rescaled seperately before combined into a well conditioned system.

The coefficient rescaling is the following:
\begin{equation*}
    \vec{j} = M \vec{y},
\end{equation*}
with
\begin{equation}
    M = i \sqrt{ \kappa_0} G_p^{-1} \mathbb{P}_p^{\Sigma H} +\frac{1}{\sqrt{\kappa_0}} G_p^{-1} \mathbb{P}_p^\Lambda.
\end{equation}
The modified block preconditioner $\hat{\mathbb{T}}_r^{(0)}$ looks like this:
 \begin{equation}
 \footnotesize
 \label{eq:lf_precond}
       \hat{\mathbb{T}}_r^{(0)} =  G_p^{-1}\left(\frac{1}{i \sqrt{\kappa_0}}\mathbb{\Theta}^{\Sigma H}_{pr} + \sqrt{\kappa_0} \mathbb{\Theta}^\Lambda_{pr}\right) G_r^{-1} T_r  G_r^{-1}\left(\Theta^\Sigma_{rp} + \Theta^{\Lambda H}_{rp}\right)  G_p^{-1} ,
 \end{equation}
 where $\mathbb{\Theta}$ is a modified version of $\Theta$. 
 \begin{align}
    \mathbb{\Theta}^{\Sigma }_{pr} =  \mathrm{P}^{\Sigma }_p G_p^{-1} N_{pr} G_r^{-1}\mathbb P^{\Lambda}_r \\
    \mathbb{\Theta}^{\Lambda H}_{pr} = \mathrm{P}^{\Lambda H}_p  G_p^{-1} N_{pr}  G_r^{-1} \mathbb  P^{\Sigma H}_r
\end{align}

The following frequency scaling holds for $\mathbb{\Theta}$ and $\Theta$:

\begin{equation}
 A^T (\Theta^\Sigma+\Theta^{\Lambda H}) A = \mathcal{O} \begin{pmatrix}
     0 & 0 & 1 \\
     0 & 0 & 1 \\
     1 & 1 & 0
 \end{pmatrix}    
\end{equation}

\begin{equation}
 A^T  (\mathbb{\Theta}^\Sigma+\mathbb{\Theta}^{\Lambda H})  A = \mathcal{O} \begin{pmatrix}
     0 & 1 & 1 \\
     0 & 1 & 1 \\
     1 & 0 & 0
 \end{pmatrix}    
\end{equation}

After applying the modified preconditioner \eqref{eq:lf_precond} and rescaling to the diagnoal and off-diagonal 2x2 blocks in \eqref{eq:precond_mt}, we find the following frequency scaling for the 2x2 blocks:

\begin{equation}
   \mathcal{O} \begin{pmatrix}
     1 & \kappa   & \kappa & \kappa & 1   & 1 \\
     1 & \kappa & \kappa & \kappa & 1   & 1 \\
     \kappa & \kappa^2 & 1 & 1 & \kappa   & \kappa \\
     \kappa & 1   & 1 & 1 & \kappa   & \kappa \\
     \kappa & 1   & 1 & 1 & \kappa & \kappa \\
     1 & \kappa   & \kappa & \kappa & \kappa^2 & 1
    \end{pmatrix}
\end{equation}

This shows that in the low-frequency limit, the quasi-Helmholtz CMP higher-order global multitrace formulation remains stable.

\section{Numerical Results}
\label{sec:num}
In this section, we present several numerical experiments to ascertain the correctness and efficiency of the preconditioning strategy for higher-order multi-trace integral equations. Results are obtained using the qH-CMP multi-trace method with orders $p=0, 1, 2$, and $3$. They are compared with various existing methods, including zeroth-order PMCHWT, multi-trace formulations that utilize Buffa-Christiansen dual functions for preconditioning, and analytic Mie series, where possible.

\subsection{Validation of solution}

\begin{figure}[hb]

    \centering
    \includegraphics[width=0.45\linewidth, trim={4cm 2cm 6cm 3cm},clip]{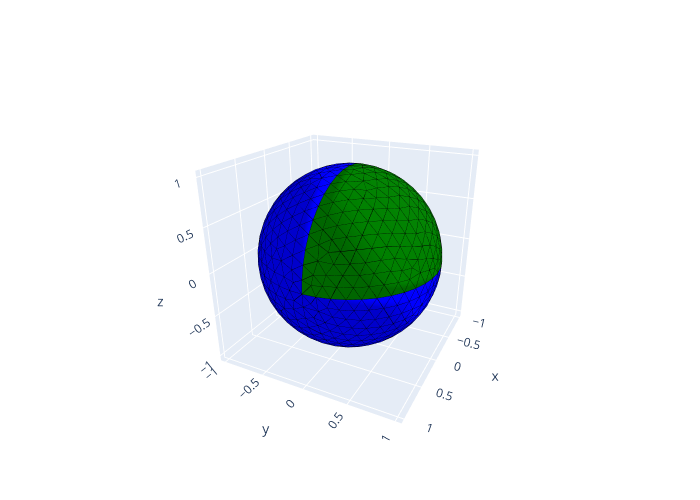}
    \includegraphics[width=0.45\linewidth, trim={4cm 1.25cm 3.5cm 3cm},clip]{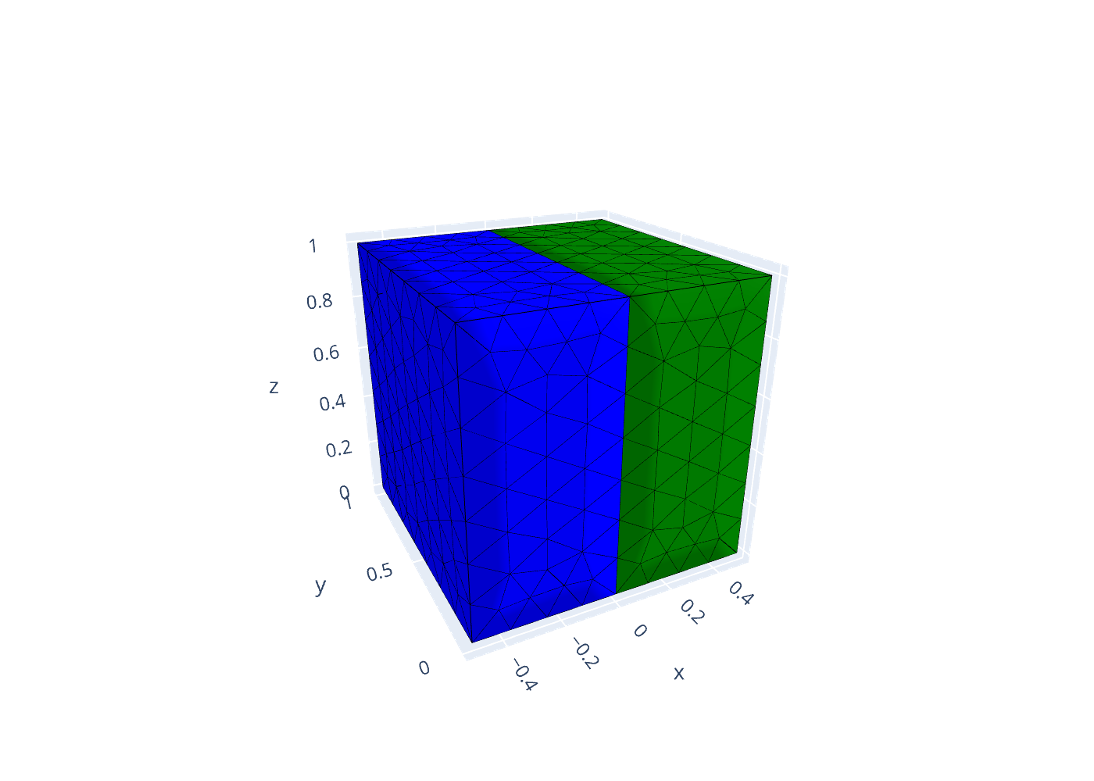}
    \caption{Geometries used to test the multi-trace formulation. A two-domain sphere. One-quarter of the sphere is treated as a separate domain. And two touching boxes that together form a unit cube.}
    \label{fig:sphere}
\end{figure}

The first validation is performed on the unit sphere with a radius of $1m$. The sphere is filled with a homogeneous material with $\varepsilon_r = 3$ and $\mu_r=0$. To test the multi-trace formulation, the unit sphere is split into two parts, one encompassing one-quarter of the sphere and the other the remaining three-quarters (see Fig.\ref{fig:sphere}), without changing the material properties of either part, so both parts together still affect an incoming excitation the same as the whole sphere. This allows comparison of the results from the multi-trace formulations with those from the Mie series solution and the PMCHWT formulation.

\paragraph*{Note} It is evident that for such a simple configuration, in practice, a multi-trace method is unnecessary. However, this simple geometry allows for comparisons with the analytical Mie series and establishes a baseline for other comparisons with more complex geometries.

\begin{figure*}
\centering

\includegraphics[width=1.0\linewidth]{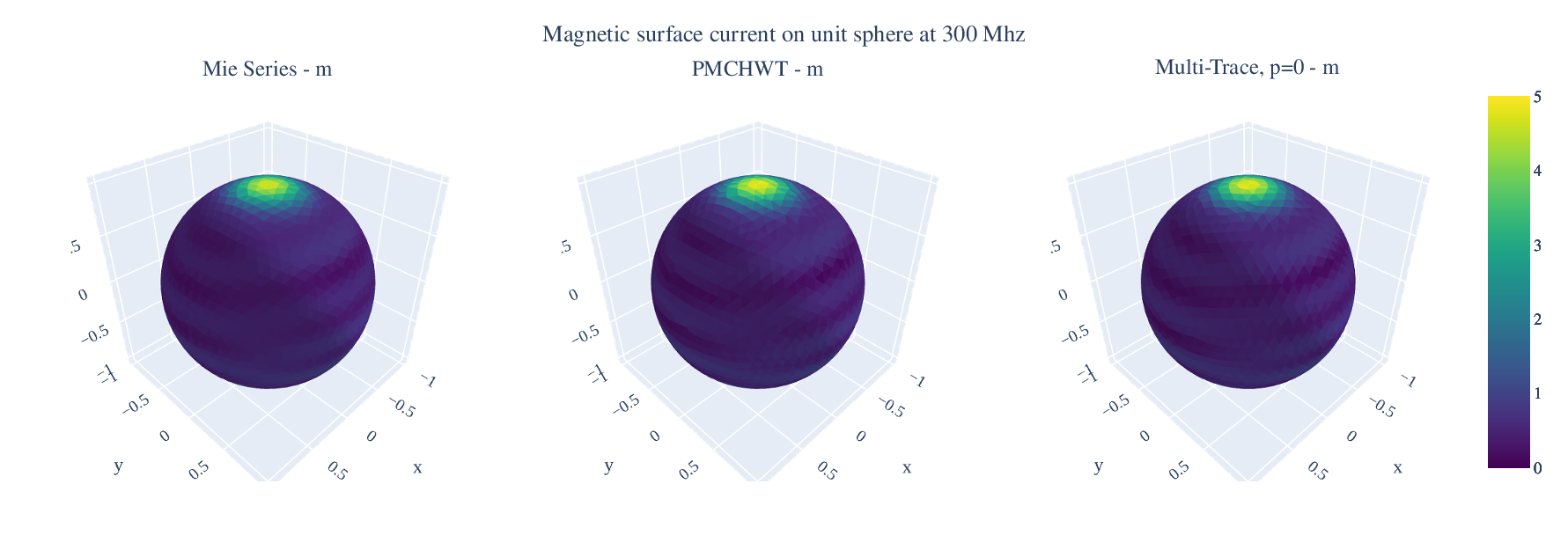}
    \caption{Magnetic surface current  ($\vec{m} = \vec{E}\times \vec{n}$) computed on the unit sphere from Mie series (left), PMCHWT (middle), and multi-trace formulation (right). The multi-trace example used the two-domain sphere (Fig. \ref{fig:sphere}). In all three examples, the meshes are the same on the sphere's outer surface. } 
    \label{fig:sphere_current}
\end{figure*}

The sphere is illuminated with an $x$-polarized plane wave propagating in the positive $z$-direction at $300MHz$. The induced surface current on the dielectric sphere can be seen in Fig. \ref{fig:sphere_current}. The surface currents obtained from the numerical schemes (PMCHWT and lowest-order multi-trace) match the analytic solution from the Mie series.

\begin{figure}
    \centering
   \includegraphics[width=0.45\linewidth]{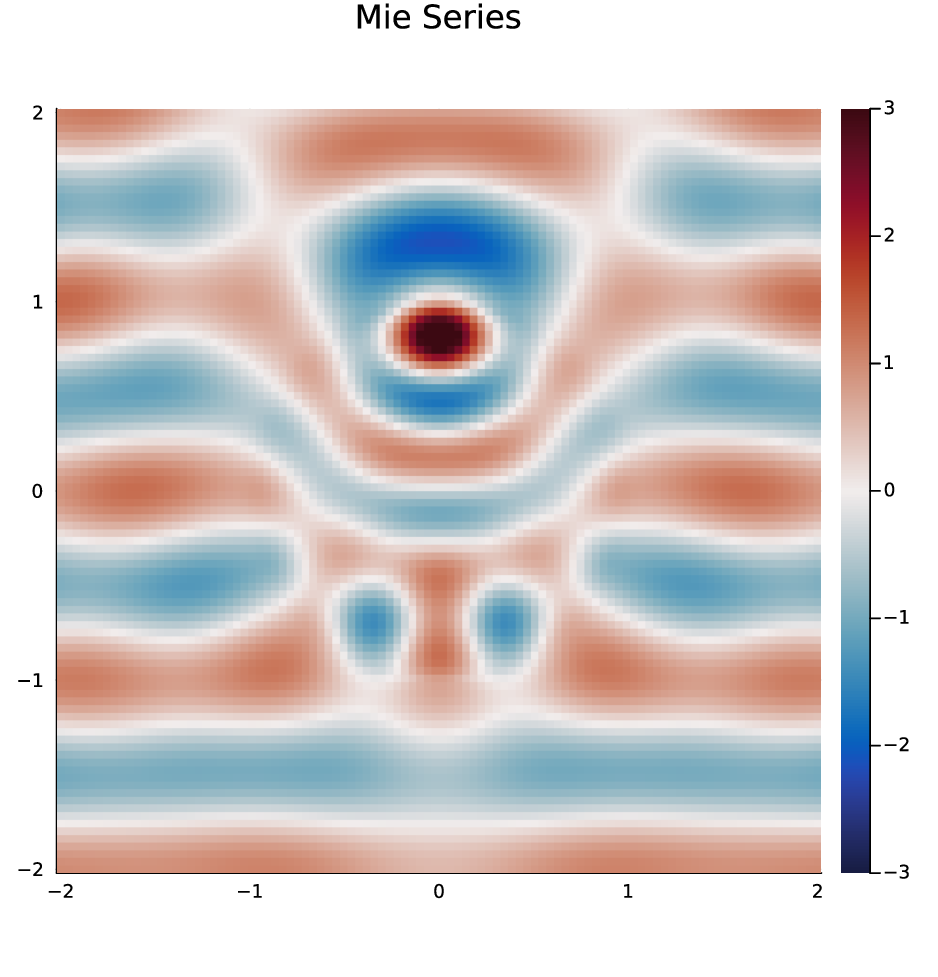}
    \includegraphics[width=0.45\linewidth]{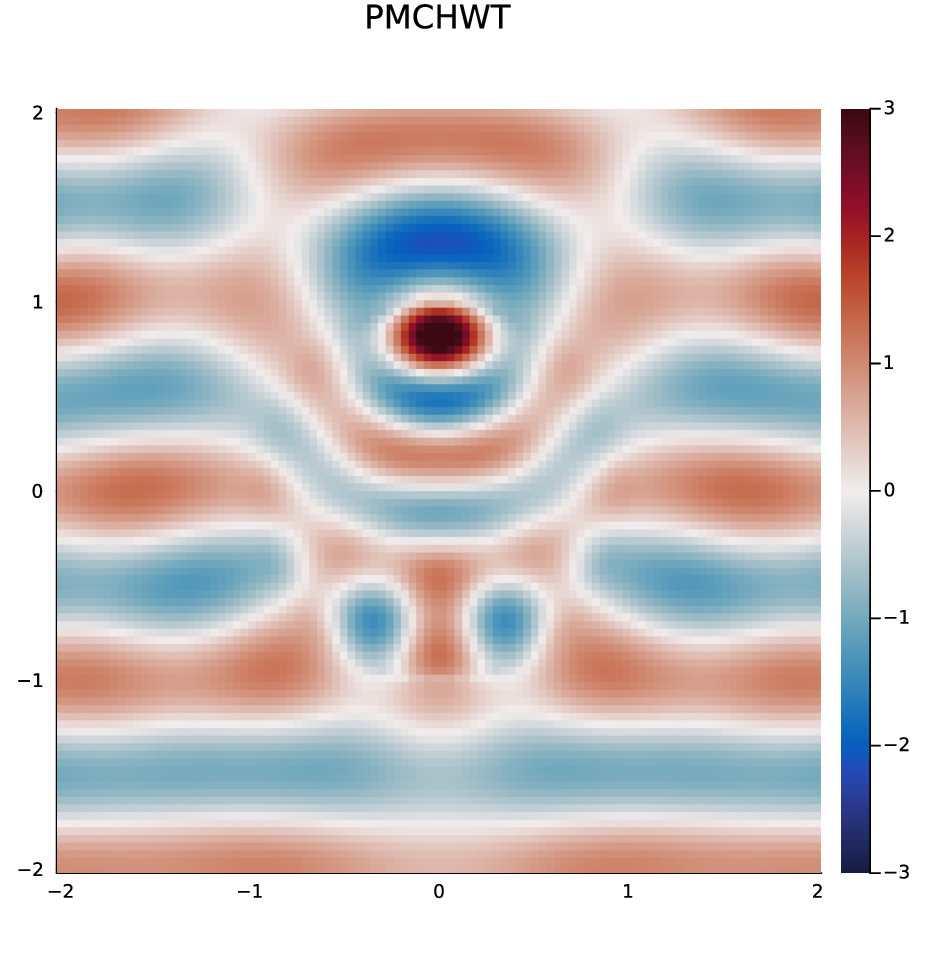}
    \vspace{1.2mm}\\
    \includegraphics[width=0.45\linewidth]{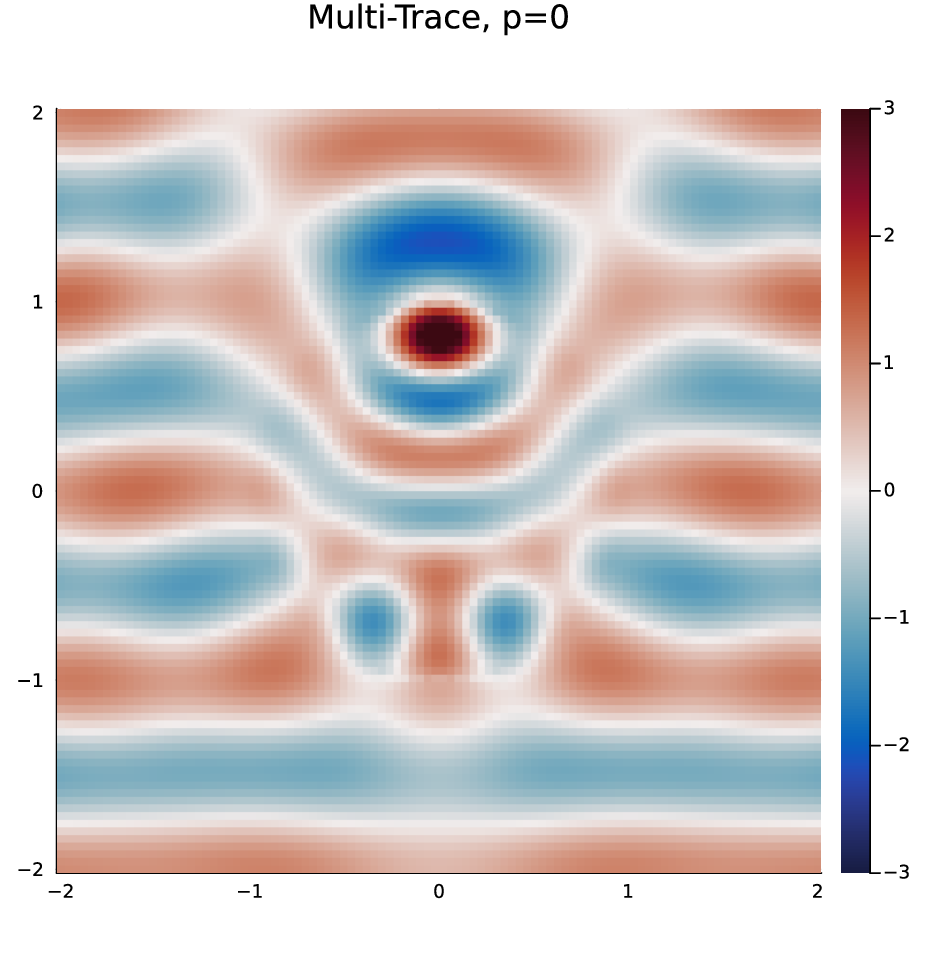}
    \includegraphics[width=0.45\linewidth]{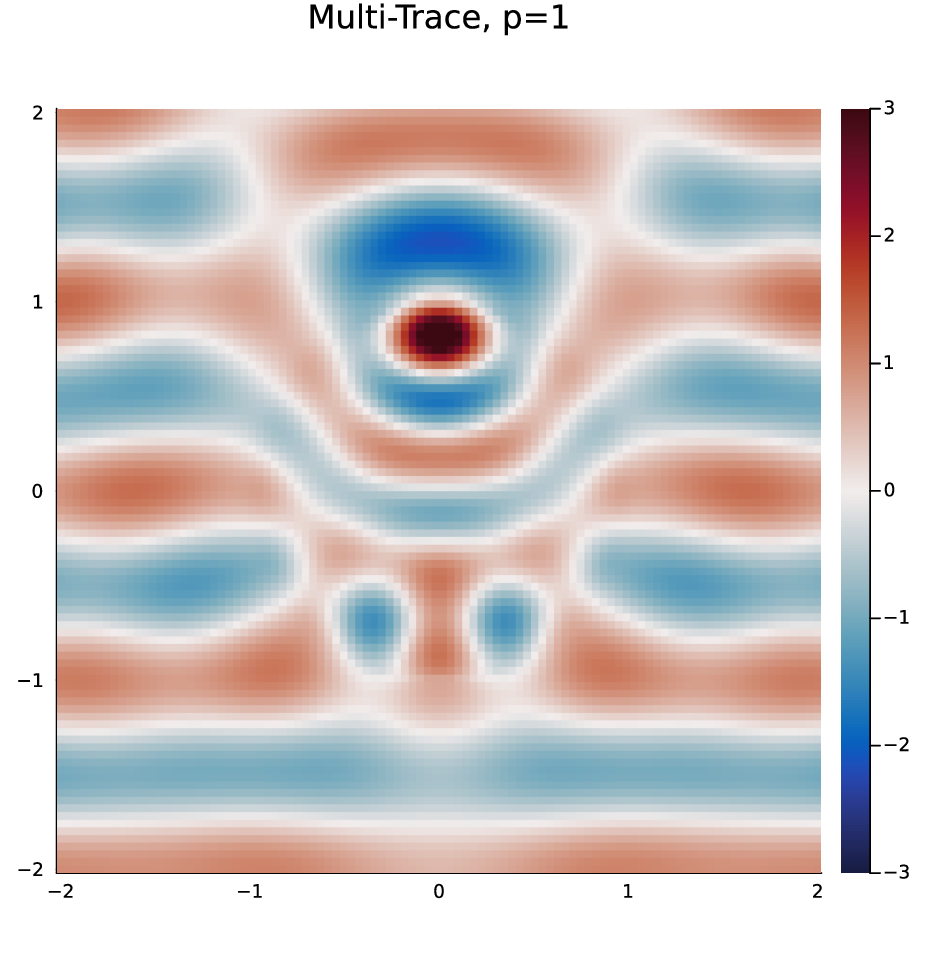}
    \vspace{1.2mm}\\
    \includegraphics[width=0.45\linewidth]{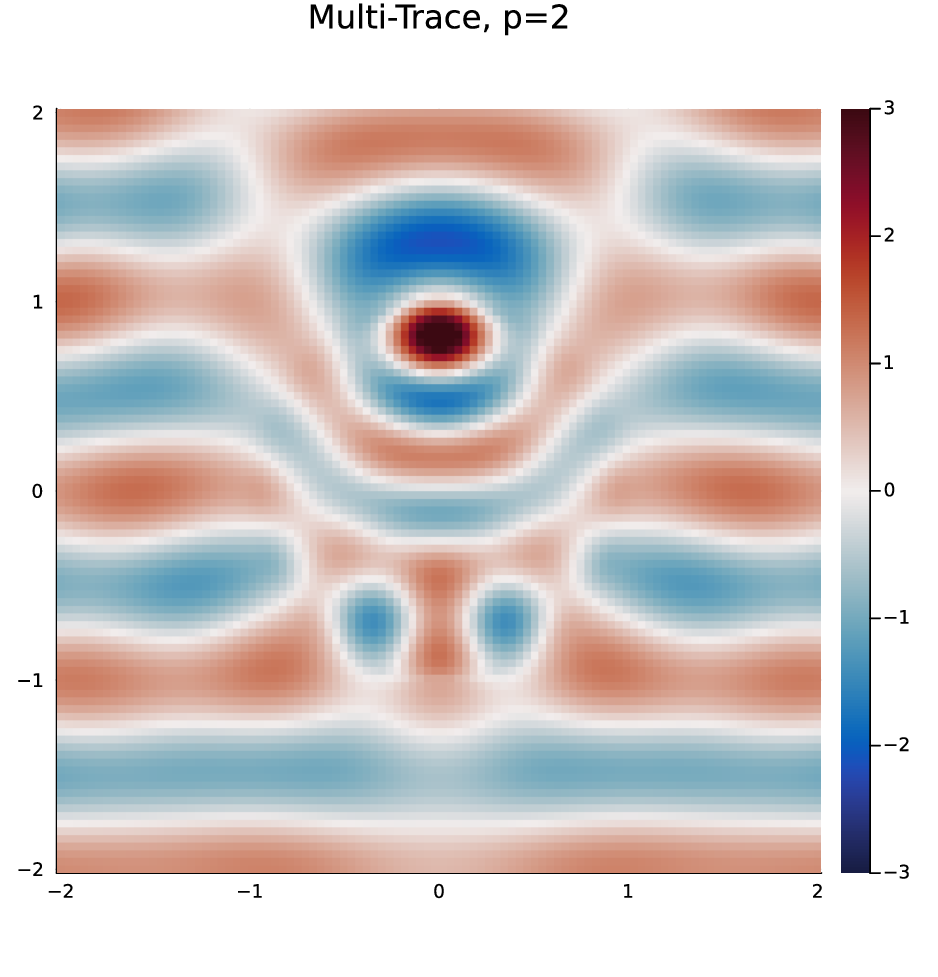}
    \includegraphics[width=0.45\linewidth]{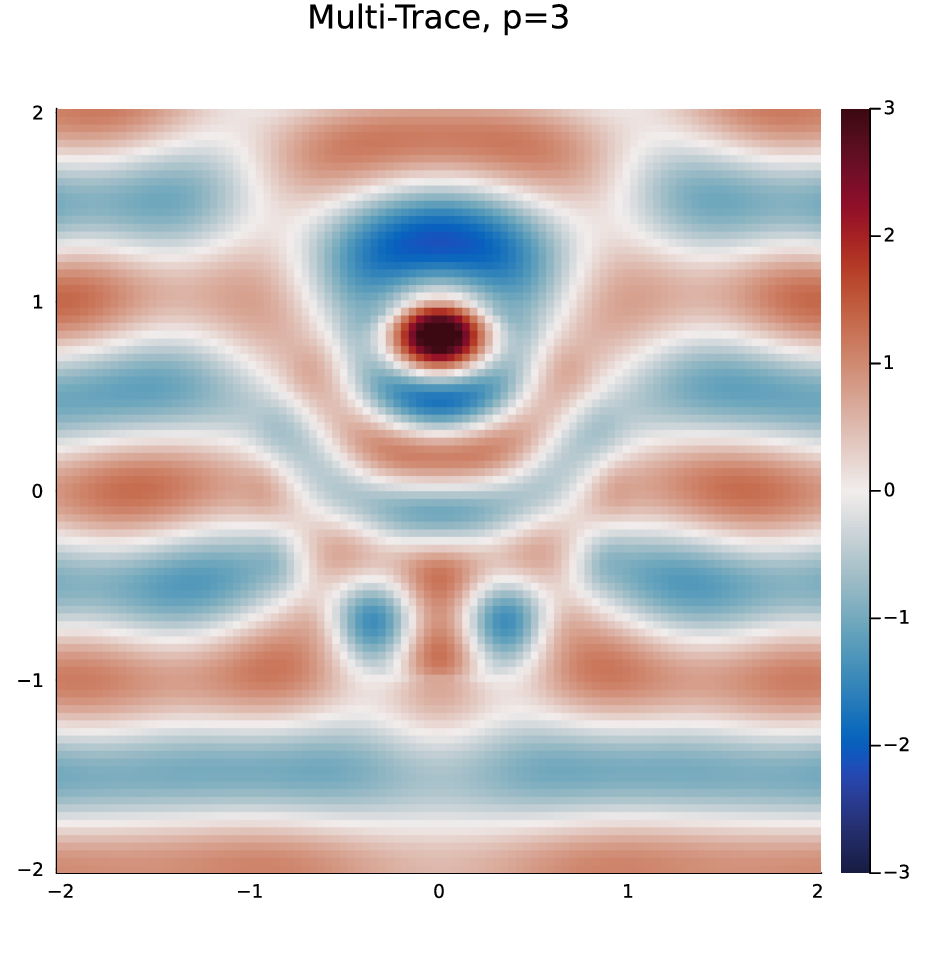}
    \vspace{0mm}\\
    \caption{Electric near field around the unit sphere in the plane ($z=0$) from Mie series (top left), PMCHWT (top right), and multi-trace of order $0$ (middle left), $1$ (middle right), $2$ (bottom left), and $3$ (bottom right).}
     \label{fig:sphere_near}
\end{figure}

The electric nearfield around the sphere can be seen in Fig. \ref{fig:sphere_near}. Again, for the analytic solution from the Mie series and for the numerical solutions from the PMCHWT and the multi-trace formulations of orders $0$, $1$, $2$, and $3$. All numerical solutions agree well with each other with roughly a $1.5\%$ relative error at the center of the sphere and $0.5\%$ relative error $1m$ away from the surface compared to the Mie series solution. 

\begin{figure}
    \centering
    \includegraphics[width=\linewidth, trim={0cm 0cm 0cm 0cm},clip]{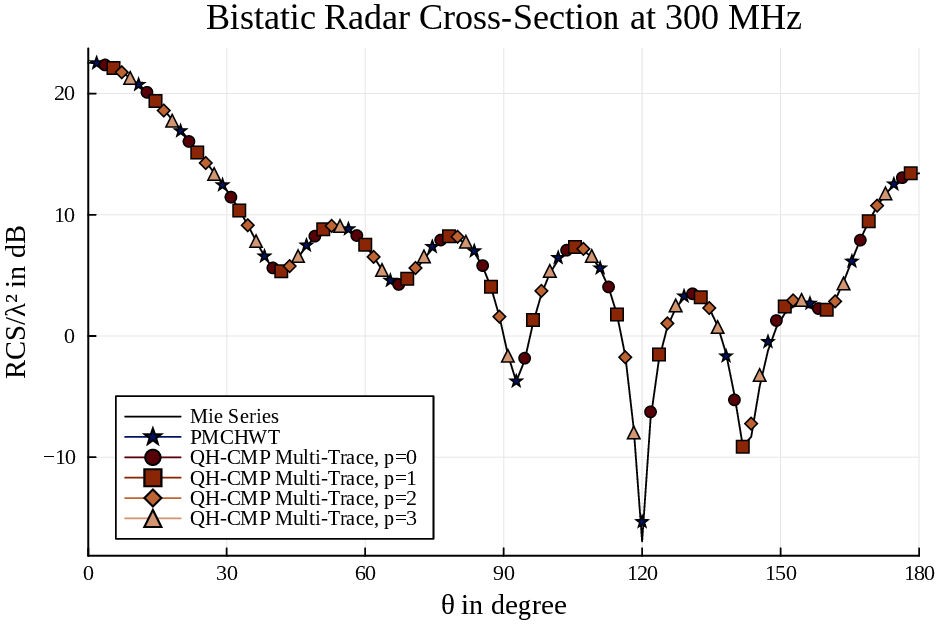}
    \caption{Bistatic radar cross-section of sphere with radius $1m$ at $300Mhz$.}
      \label{fig:sphere_far}
\end{figure}

The bistatic radar cross section of the unit sphere illuminated by a $300MHz$ plane wave is visualized in Fig. \ref{fig:sphere_far}. Once again, one can see that all the solutions (PMCHWT and multi-trace of order $p=0,1,2,3$) match the reference solution from the Mie series.

The root-mean-square (RMS) error relative to the Mie series is $6.44\%$ in the far field for the PMCHWT and between $5.85\%$ and $7.28\%$ for the multi-trace formulation, with the zeroth-order solution being the most accurate. 

A reason the higher-order basis function does not yield lower error is that, on the sphere, the primary source of error is the geometric error introduced by using flat triangles to discretize the sphere's curved geometry. 

\subsection{Convegence}

Before looking at the impact of the higher-order basis function, the convergence of in the energy norm is studied for the case of two touching boxes (cf. Fig. \ref{fig:sphere}). For that, the discrete energy norm is computed relative to a very fine discretization ($h=0.05$) with the following formula: 

\begin{equation}
    \label{eq:energynorm}
     \Vert u_h-u_\circ \Vert_S = \sqrt{u_h^* S_{ff} u_h - 2 \operatorname{Re}(u_h^* S_{ff_\circ} u_\circ) + u_\circ^* S_{f_\circ f_\circ} u_\circ},
\end{equation}
where $f$ is the basis for the numerical solution and $f_\circ$ is the basis for the reference solution and with  
\begin{align}
    \left( \mat{S}_{ff} \right)_{m,n} := &  \iint_{\Gamma \times \Gamma} \frac{f_m(x) \cdot f_n(y)}{4 \pi |x-y|} dy dx  + \notag \\
    &  \iint_{\Gamma \times \Gamma} \frac{\operatorname{div}_{\Gamma} f_m(x) \operatorname{div}_{\Gamma} f_n(y)}{4 \pi |x-y|} dy dx.
\end{align}

\begin{figure}
    \centering
    \includegraphics[width=\linewidth, trim={0cm 0cm 0cm 0cm},clip]{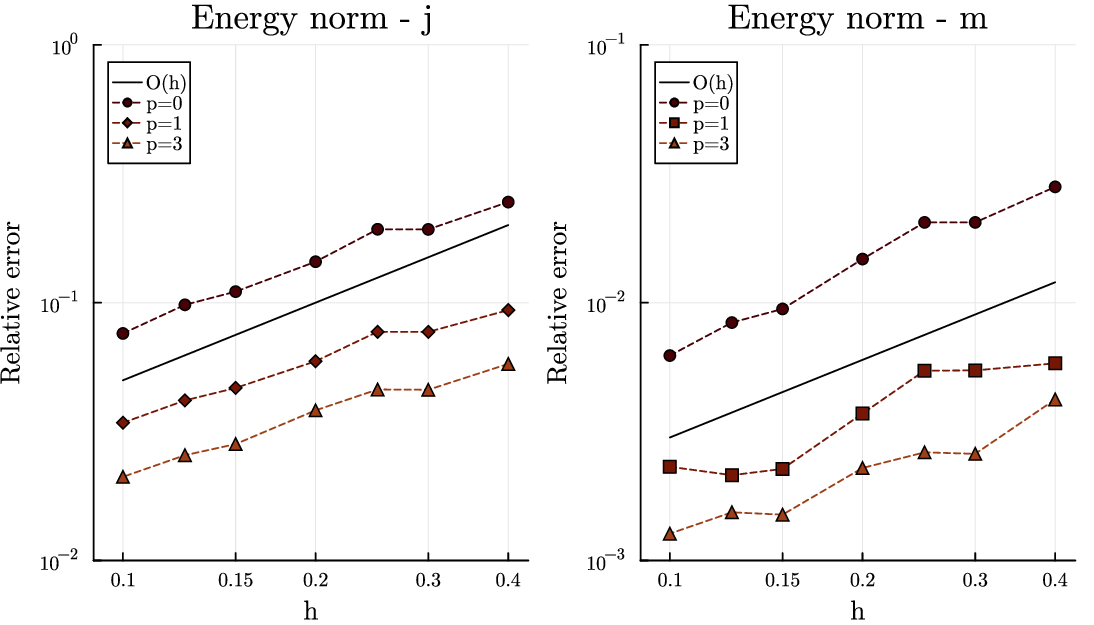}
    \caption{Error in the energy norm relative to the solution on fine mesh ($h=0.05$).}
      \label{fig:energynorm}
\end{figure}

Fig. \ref{fig:energynorm} shows the convergence of the energynorm for the multi-trace formulation of order $p=0,1,2$ for two touching boxes. All methods converge at roughly the same rate of 1. For the higher-order methods, the convergence is constrained by the geometrically sharp features of the boxes.
Nevertheless, the higher-order methods generally achieve lower errors, providing comparable or even improved accuracy on coarser meshes. As a result, they require fewer degrees of freedom and produce smaller system sizes.

\subsection{Impact of higher-order basis}

Next, we investigate the impact of using higher-order basis functions in the multi-trace formulation. Two touching boxes (Fig. \ref{fig:sphere}) are used to avoid limited accuracy due to geometric errors. The flat surfaces of the boxes can be represented exactly with flat triangular patches. Thus, unlike in the previous example of the sphere, there will be no geometric approximation error. However, there is no longer an analytic reference solution in the form of the Mie series; instead, we used a high-order solution ($p=3$) on a fine mesh as the reference solution. 

\begin{figure}
    \centering
    \includegraphics[width=\linewidth, trim={0cm 0cm 0cm 0cm},clip]{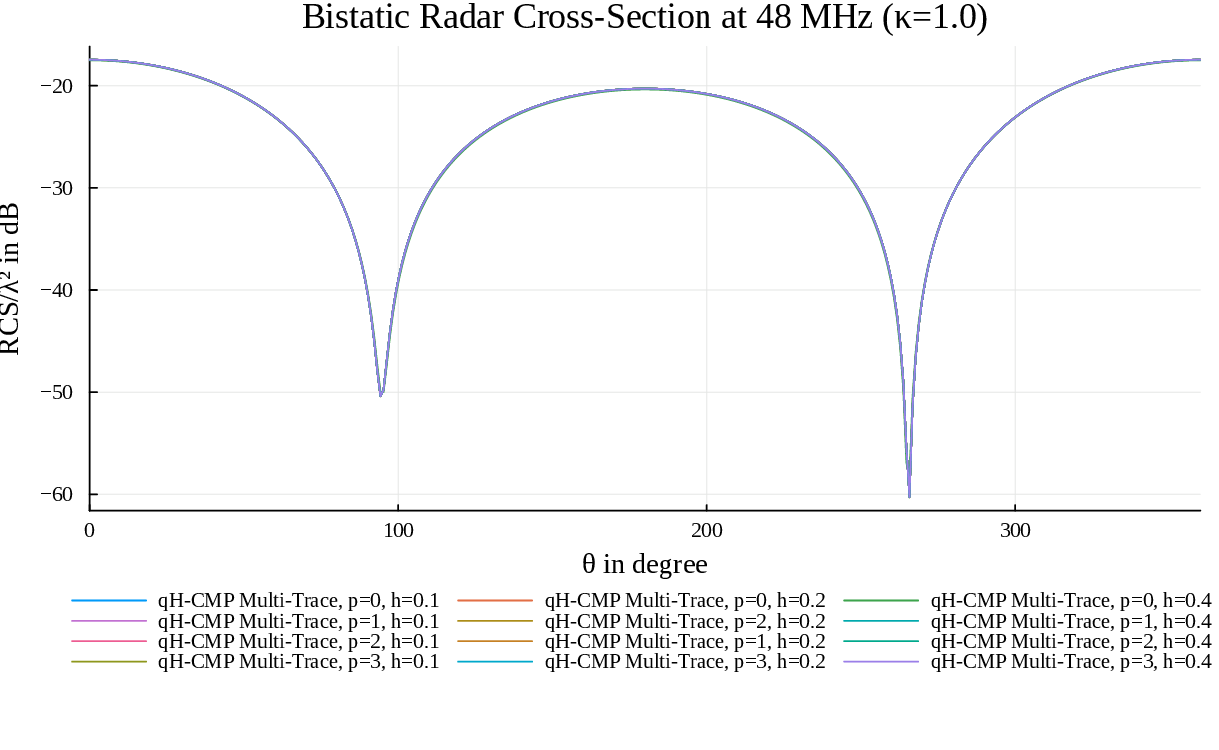}
    \caption{Bistatic radar cross-section of two touching boxes at $48Mhz$ computed from the numerical solution obtained with different settings (varying mesh size $h$ and order $p$ of basis functions). }
    \label{fig:boxes_farfield}
\end{figure}

\begin{figure}
    \centering
    \includegraphics[width=\linewidth, trim={0cm 0cm 0cm 0cm},clip]{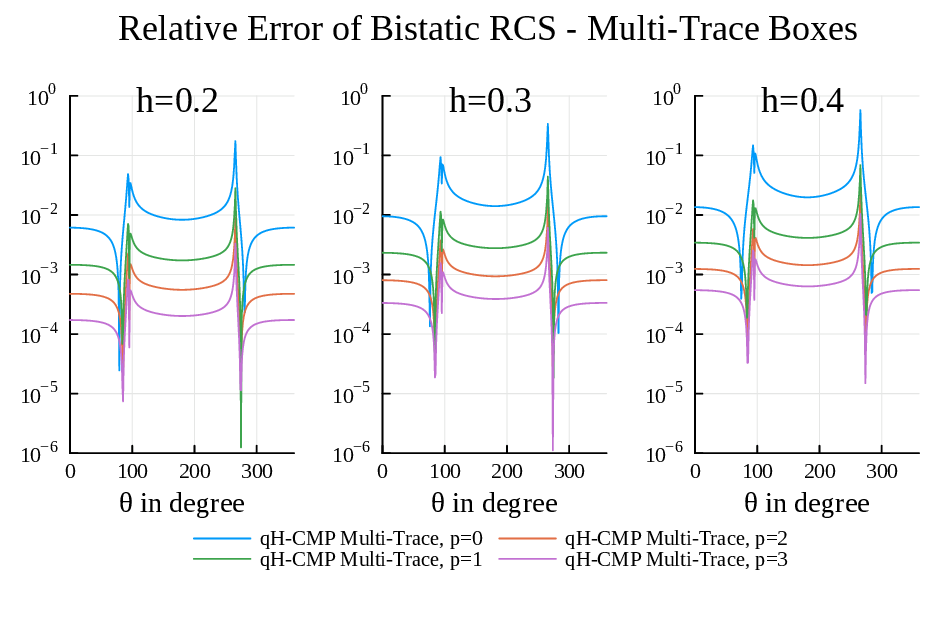}
    \caption{Relative error of bistatic radar cross-section of two touching boxes at $48Mhz$ at different mesh sizes. }
    \label{fig:rcs_error_h}
\end{figure}

\begin{figure}
    \centering
    \includegraphics[width=\linewidth, trim={0cm 0cm 0cm 0cm},clip]{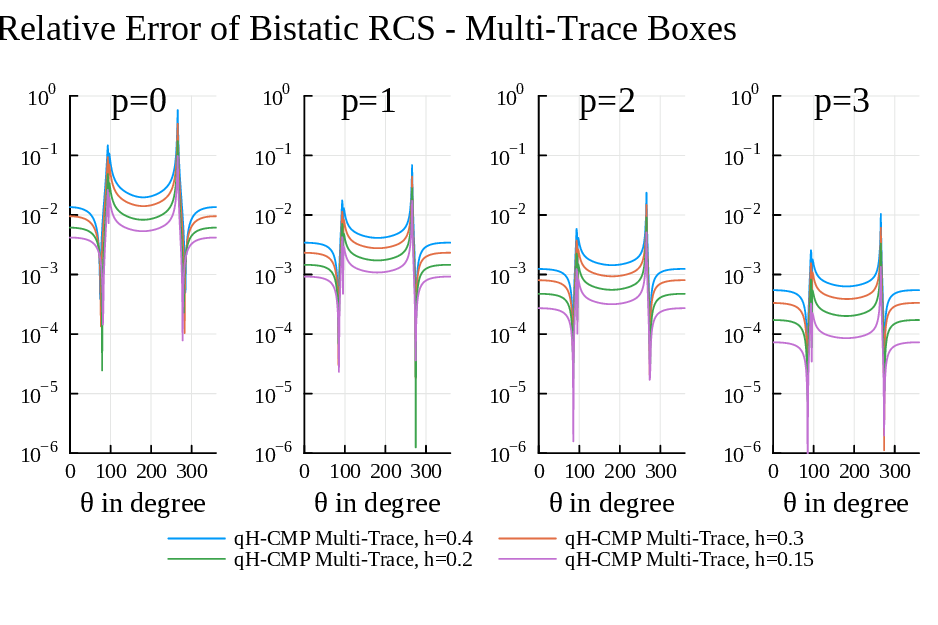}
    \caption{Relative error of bistatic radar cross-section of two touching boxes at $48Mhz$ at different orders of basis functions. }
    \label{fig:rcs_error_p}
\end{figure}

A similar excitation is used; an x-polarized plane wave traveling in the z-direction at $48MHz~(\kappa_0=1)$ illuminates the composite structure. One box has a relative permittivity of $\varepsilon_1=6$, while the other's permittivity is higher at $\varepsilon_2=10$. The radar cross-section (RCS) reflects this asymmetry of the geometry (Fig. \ref{fig:boxes_farfield}). In the same figure, the RCSs computed with different settings appear identical. However, on closer inspection, the relative error shows that RCSs computed from the higher-order solutions are significantly closer to the reference solution (Fig. \ref{fig:rcs_error_h} and Fig. \ref{fig:rcs_error_p}). Although the error decreases for all orders $p$ as the mesh size $h$ is reduced, in absolute value, the higher-order method produces a much more accurate solution on a coarse mesh, even exceeding the accuracy of the lower-order methods on a finer mesh. This can be seen clearly in Fig. \ref{fig:rcs_error_p}, when the relative errors are grouped by order. 

The gain in accuracy with higher-order basis functions comes at a cost. The higher-order spaces have more degrees of freedom on the same mesh and thus lead to considerably larger, computationally more expensive linear systems. Table \ref{tab:dof} provides an overview of the number of degrees of freedom for different order basis functions on the same mesh. Going from zeroth-order to first-order basis functions increases the number of unknowns by a factor of approximately three and a half, to second-order by a factor of seven, and to third-order by a factor of more than twelve. This increase in computational cost is also reflected in the time it takes to assemble and solve the linear system. The breakdown of the total time for the different orders and mesh is listed in Table \ref{tab:time}. The timings were obtained on a server with 2 AMD EPYC 7763 Processors and 4 TB of RAM. The code was limited to using 48 threads in parallel.

\begin{table}
    \centering
    \caption{Comparison of the number of DoFs in the higher order multi-trace formulation for different orders $p$ and different mesh sizes $h$.}
    \begin{tabular}{c|r|r|r|r|r|}
              & $h=0.40$ & $h=0.30$ & $h=0.20$ & $h=0.15$ & $h=0.10$   \\ \hline
        $p=0$ & $  648$ & $1,056$ & $1,716$     & $3,048$     & $6,012$ \\
        $p=1$ & $2,260$ & $3,520$ & $5,720$     & $10,160$     & $20,040$ \\
        $p=2$ & $4,536$ & $7,392$ & $12,012$     & $21,336$     & $42,084$ \\
        $p=3$ & $7,776$ & $12,672$ & $20,592$     & $36,576$     & $72,144$ \\
    \end{tabular}
    \label{tab:dof}
\end{table}

\begin{table}
    \centering
    \caption{Total time - assembling and solving the linear system.}
    \begin{tabular}{c|r|r|r|r|r|}
              & $h=0.40$ & $h=0.30$ & $h=0.20$ & $h=0.15$ & $h=0.10$   \\ \hline
        $p=0$ & $  42.65s$ & $77.23s$ & $102.56s$     & $180.96s$     & $383.73s$ \\
        $p=1$ & $180.96s$ & $235.50s$ & $361.72s$     & $625.35s$     & $1265.82s$ \\
        $p=2$ & $484.00s$ & $637.97s$ & $936.36s$     & $1627.08s$     & $3155.46s$ \\
        $p=3$ & $1298.18s$ & $1536.45s$ & $2524.89s$     & $4302.57s$     & $8306.44s$ \\
    \end{tabular}
    \label{tab:time}
\end{table}

However, going back to the observation that the error with the higher-order basis function is much lower. One can see that with higher orders, one gets better or the same accuracy for roughly the same number of unknowns, or in other words, for the same computational effort.

\subsection{Conditioning and iterative solution}

Using a higher-order basis function only is attractive if the resulting linear system can also be solved efficiently. Due to the large number of unknowns involved in higher-order discretization, iterative solvers such as GMRES \cite{saad_gmres_1986} are the only viable option for efficiently solving the large linear system. For GMRES to converge quickly, the matrix needs to be well-conditioned, i.e., have a condition number close to unity, which ideally is independent of parameters such as the mesh size $h$ or the order of the basis function $p$ used in the discretization. In this section, the conditioning of the qH-CMP multi-trace will be investigated.

\subsubsection{Higher order projectors}

First, we look at the iterative computation of the higher-order projectors. As shown earlier in section \ref{sec:qh}, the action of the (higher-order) qH-projectors can be computed by iteratively solving a saddlepoint problem instead of computing a pseudo inverse. While the pseudo-inverse only needs to be calculated once to get an explicit expression for the projectors, the saddlepoint has to be solved every time the projectors act on a coefficient vector (i.e., in every iteration of GMRES when solving the linear system). Since the preconditioned multi-trace formulation \eqref{eq:precond_mt} involves multiple projectors, and their action must be computed in every iteration of the solver, the computation of the projectors' action should be efficient.

Table \ref{tab:saddlepoint_iter} shows the maximum number of iterations to solve the saddle point system up to the given tolerance ($10^{-8}$) when the projector's action is computed during the iterative solving process of the multi-trace system. When the preconditioning strategy from \cite{powell_optimal_2003} is employed, the iteration count is very low ($< 10$) and is independent of the number of unknowns and the order of the basis functions involved.

The setup time for the projectors (assembly of the involved sparse matrices and the Cholesky factorization of the saddle point problem) is less than $5\%$  for third-order problems and less than $3\%$ for zeroth-order problems of the total solution time (see Table \ref{tab:time}).

\begin{table}
    \centering
    \caption{Maximum number of iterations to compute the action of the projectors iteratively with a relative tolerance of $10^{-8}$ during the iterative solving of the multi-trace system}
    \begin{tabular}{c|r|r|r|r|r|}
              & $h=0.40$ & $h=0.30$ & $h=0.20$ & $h=0.15$ & $h=0.10$   \\ \hline
        $p=0$ &  $8 $ & $8 $ & $8$ & $8$ & $8$ \\
        $p=1$ &  $8 $ & $8 $ & $8$ & $8$ & $8$ \\
        $p=2$ &  $7 $ & $8 $ & $8$ & $8$ & $8$ \\
        $p=3$ &  $8 $ & $8 $ & $9$ & $9$ & $9$ \\
       \end{tabular}
    \label{tab:saddlepoint_iter}
\end{table}

\subsubsection{Higher-order multi-trace formulation}

In this part, the performance of GMRES and the effect of quasi-Helmholtz multiplicative Calder\'on preconditioning for the higher-order multi-trace formulation are examined.

\begin{figure}
    \centering
    \includegraphics[width=1\linewidth,trim={0cm 1.35cm 0cm 0cm},clip]{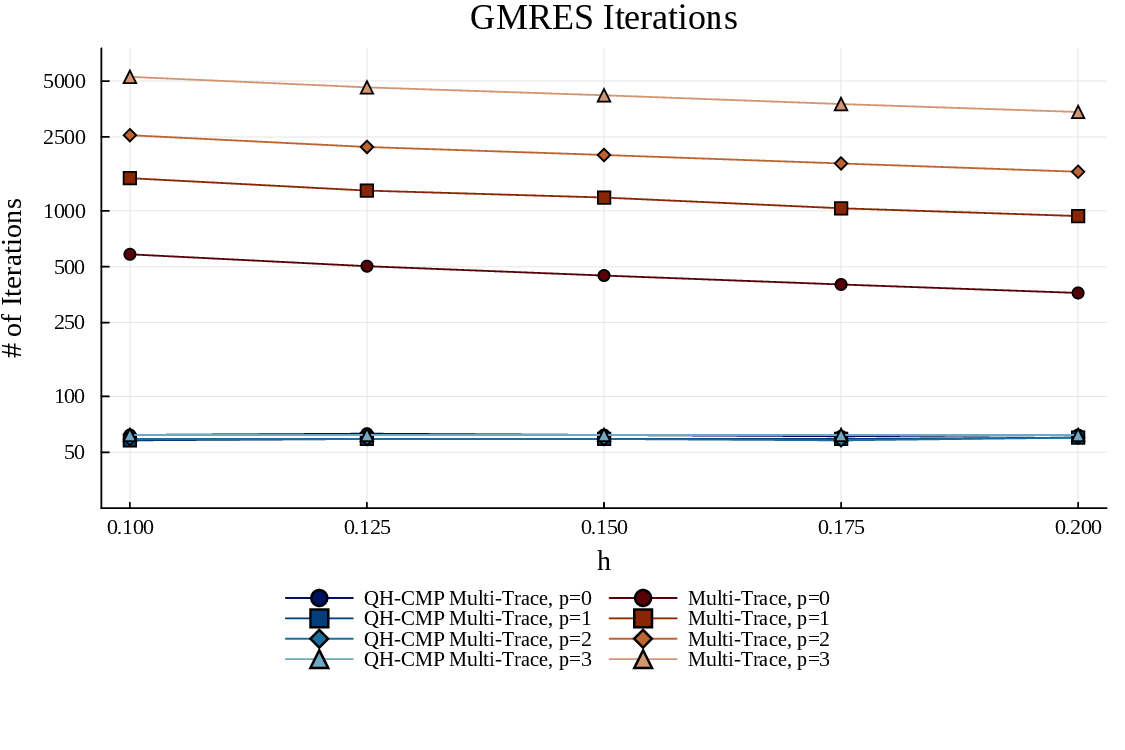}
    \caption{Number of iterations it takes GMRES to converge to a relative tolerance of $10^{-8}$ for preconditioned and regular higher-order multi-trace formulations of order $p=0,1,2$ and $3$. }
    \label{fig:gmres}
\end{figure}
\begin{figure}
    \centering
    \includegraphics[width=1\linewidth,trim={0cm 1.35cm 0cm 0cm},clip]{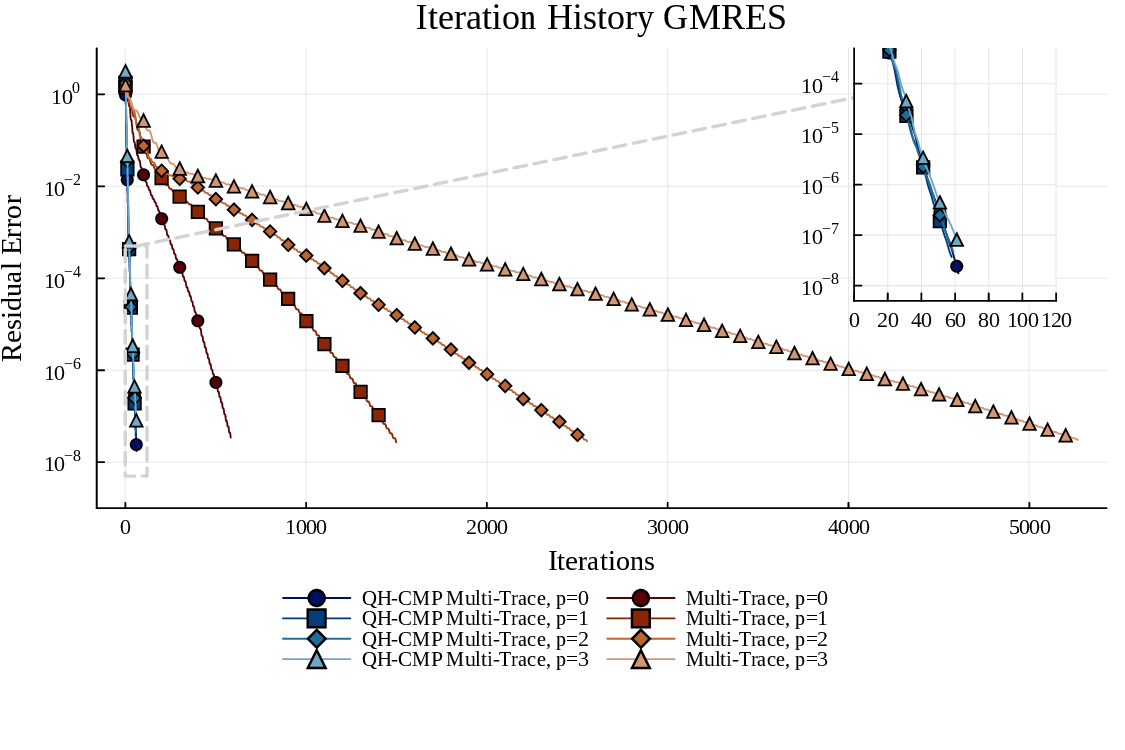}
    \caption{Convergence history of GMRES at $h=0.1$.}
    \label{fig:history}
\end{figure}

To that end, the regular and preconditioned multi-trace systems are computed for different mesh sizes and orders of basis functions. In all numerical experiments, a relative tolerance of $ 10^ {- 8} $ was used in GMRES, and no restarts were performed during the iteration. Fig. \ref{fig:gmres} shows that the number of iterations for GMRES solving the multitrace formulation for an incident plane wave at $48MHz$ on two boxes with $\varepsilon_1=2$ and $\varepsilon_2=3$.
The system's conditioning is directly reflected in the number of GMRES iterations required for the solver to converge. The iteration count is significantly higher for the regular system than for the preconditioned system and is influenced by the order of the basis functions. In contrast, in the preconditioned system, the number of iterations is independent of the mesh size and order of the basis functions. In this example, at moderate frequency and relatively low contrast, GMRES converges in approximately $60$ iterations when the quasi-Helmholtz Calder\'{o}n multiplicative preconditioner is applied to the system. For the regular system discretized with third-order basis functions, GMRES needs over $5,000$ iterations to converge. Fig. \ref{fig:history} shows the iteration history for $h=0.1$. The drastic slowdown in the convergence of the regular system, when higher-order basis functions are employed, is evident in the convergence history.

Table \ref{tab:projector_iter} shows the GMRES iteration for two boxes with higher permittivity ($\varepsilon=6$ and $\varepsilon_2=10$). The GMRES iteration count increases overall compared to the previous example. There is a slight variation between different mesh sizes and the order of basis functions, but the overall number of iterations remains relatively low.

\begin{table}
    \centering
    \caption{Number of iterations to solve the preconditioned multi-trace system iteratively with a relative tolerance of $10^{-8}$}
    \begin{tabular}{c|r|r|r|r|r|}
              & $h=0.40$ & $h=0.30$ & $h=0.20$ & $h=0.15$ & $h=0.10$   \\ \hline
        $p=0$ & $ 129$ & $ 141$ & $ 152$ & $ 148$ & $ 148$ \\
        $p=1$ & $ 136$ & $ 143$ & $ 142$ & $ 139$ & $ 135$\\
        $p=2$ & $ 132$ & $ 138$ & $ 135$ & $ 133$ & $ 132$\\
        $p=3$ & $ 138$ & $ 142$ & $ 138$ & $ 136$ &
        $ 136$\\
       \end{tabular}
    \label{tab:projector_iter}
\end{table}

\subsection{Very low frequency}

\begin{figure}[hb]
    \centering
    \includegraphics[width=0.45\linewidth, trim={0cm 0cm 0cm 0cm},clip]{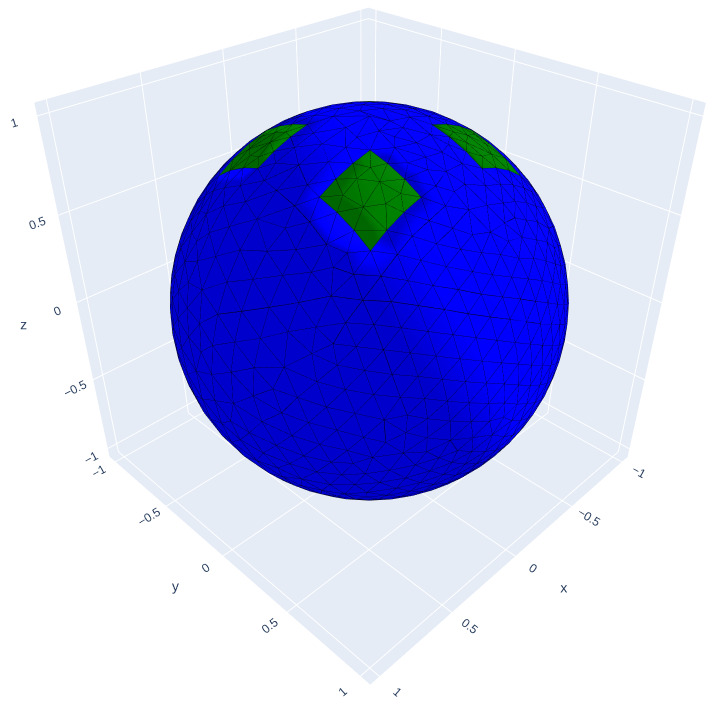}
    \includegraphics[width=0.45\linewidth, trim={0cm 0cm 0cm 0cm},clip]{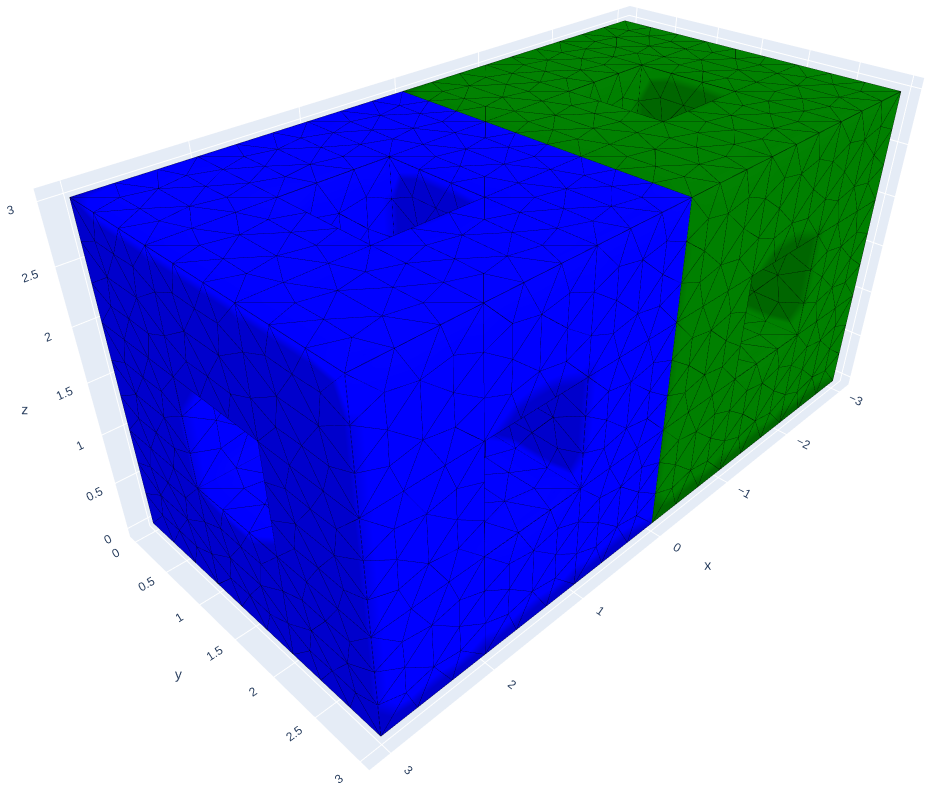}
    \caption{Multiply-connected geometries used to test the multi-trace formulation at very low frequency. A sphere with four pillars in the center that are treated as a separate domain. And two touching cubes with holes in all faces. The edge length of the cubes is $3m$ and the edge length of the holes is $1m$}
    \label{fig:lfgeometry}
\end{figure}

Last, we study two more geometries to analyze the low-frequency behavior of the multi-trace formulation. The first geometry is a sphere with four rectangular blocks removed from its center. The second one consists of two adjoining cubes with rectangular hole cuts through all faces (Fig. \ref{fig:lfgeometry}).

The following plane wave excitation is used, which has neither the direction nor the polarization aligned with a coordinate axis. 
\begin{equation}
    \label{eq:lfpw}
    \vec{e}^{inc}(\vec{r}) = \left(-\frac{1}{\sqrt{2}},\frac{1}{\sqrt{2}},0\right)^{T} \exp \left( -\frac{i \kappa_0}{\sqrt{3}} (x+y+z) \right)
\end{equation}

Both parts of the geometry are filled with the same material ($\varepsilon_r = 3$), allowing comparison with the Mie series and PMCHWT using a geometry obtained by fusing the two parts. The low-frequency stabilized version of the PMCHWT \cite{beghein_low-frequency_2017} is used, as a very low wavenumber ($\kappa_0=10^{-40}$) is employed in the experiments.

\begin{figure}
    \centering
    \includegraphics[width=\linewidth, trim={0cm 1cm 0cm 0cm},clip]{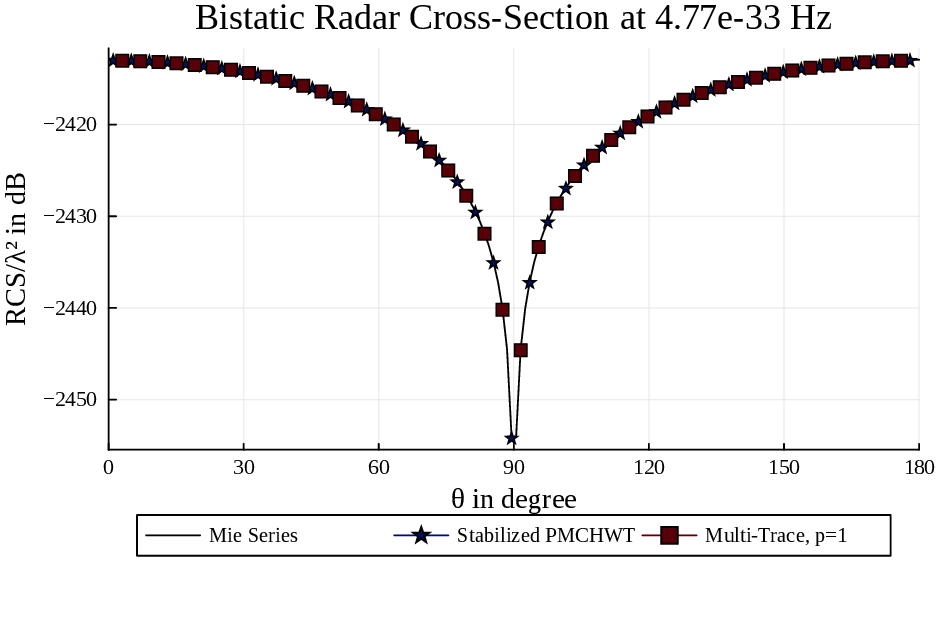}
    \caption{Bistatic radar cross-section of sphere with radius $1m$ and filled with $\varepsilon_r=3$ at $4.77\times10^{-33}Hz (\kappa=10^{-40})$. The RCS is computed with the Mie series, the low-frequency stabilized PMCHWT, and the modified multi-trace formulation of order 1.   }
      \label{fig:lfrcs_sphere}
\end{figure}

The sphere in Fig. \ref{fig:lfgeometry} is illuminated with the plane wave \eqref{eq:lfpw} at $4.77\times 10^{-33}Hz $  ($\kappa_0=10^{-40}$). The resulting farfield can be seen in Fig. \ref{fig:lfrcs_sphere}. The RCS obtained from the numerical schemes (low-frequency stabilized PMCHWT \cite{beghein_low-frequency_2017} and first-order multi-trace [\eqref{eq:precond_mt} using \eqref{eq:lf_precond}]) closely match the analytic solution from the Mie series.

\begin{figure}
    \centering
    \includegraphics[width=\linewidth, trim={0cm 1cm 0cm 0cm},clip]{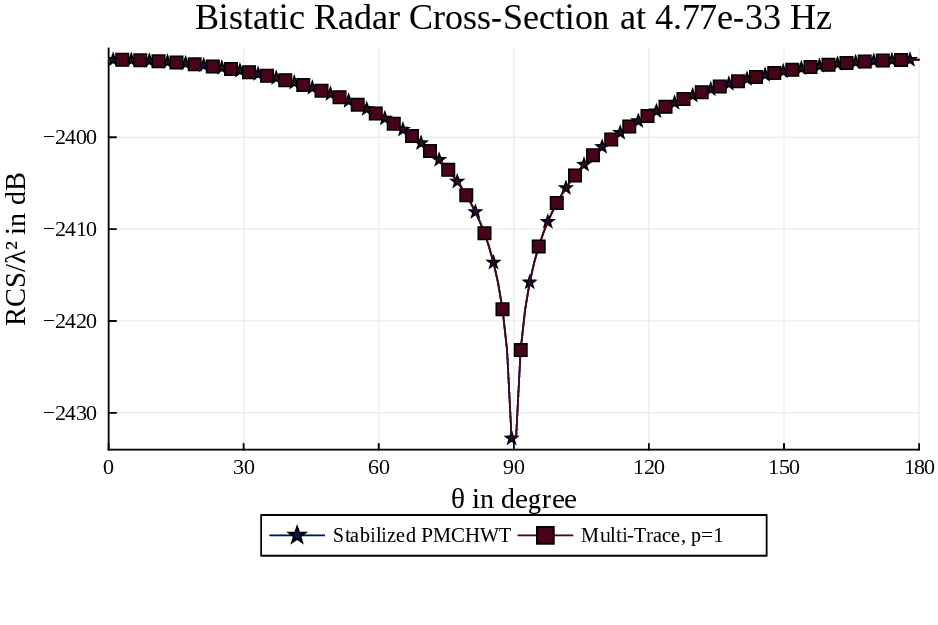}
    \caption{Bistatic radar cross-section of holed cubes with edge length $3m$ and filled with $\varepsilon_r=3$ at $4.77\times10^{-33}Hz (\kappa=10^{-40})$. The RCS is computed with the low-frequency stabilized PMCHWT and the modified multi-trace formulation of order 1.   }
      \label{fig:lfrcs_box}
\end{figure}

Fig. \ref{fig:lfrcs_box} shows the RCS for the holed box geometry in Fig. \ref{fig:lfgeometry}. Again, the PMCHWT and the multi-trace solution match. 

Table \ref{tab:globalloop_iter} lists the number of iterations to solve the scattering problem in the geometries with global loops present for different mesh resolutions. The variation in iteration count is low, indicating that the mesh size does not affect the solver's convergence. 

\begin{table}
    \centering
       \caption{Number of iterations to solve multiply connected geometry with zeroth-order multi-trace formulation at very low frequency ($\kappa_0=10^{-40}$)}
    \begin{tabular}{c|l|c|c|l|c|}
         &  h & Iterations &  & h  & Iterations \\ \hline
       \multirow{4}{*}{\rotatebox[origin=c]{90}{Sphere}}  &  0.4&  148&    \multirow{4}{*}{\rotatebox[origin=c]{90}{Boxes}} &  0.5& 106\\
         &  0.3&  163&  &  0.4& 107\\
         &  0.2&  143&  &  0.3& 118\\
         &  0.1& 155&  & 0.2& 115\\
    \end{tabular}
 
    \label{tab:globalloop_iter}
\end{table}

\section{Conclusion}
This work employed higher-order quasi-Helmholtz projectors to construct a Calder\'on Multiplicative Preconditioner for the higher-order multi-trace formulation that does not require dual basis functions. A new method for iteratively computing the action of higher-order projectors was introduced. Without the expensive calculation of a pseudo-inverse, the higher-order projectors can be used in conjunction with standard GMRES to build a fast solver for higher-order multi-trace formulations.

Numerical results show that the number of iterations for the preconditioned system is independent of the mesh size and the order of the basis function used in the discretization. With higher-order basis functions, the error can be significantly reduced; however, this comes at the expense of increased time and computational resources.

Higher-order projectors enable the stabilization of higher-order multi-trace methods at very low frequencies. Results show that, for multiply connected domains, only small modifications to the higher-order multi-trace formulations are required to render them stable at very low frequencies.

For improved results on smooth geometries, a better geometric approximation, such as curved triangles or NURBS, could further improve the formulation. When using higher-order basis functions, the geometric error quickly becomes the dominant error in the approximation, obscuring the benefits gained from higher-order basis functions.

%

\ifCLASSOPTIONcaptionsoff
  \newpage
\fi



\bibliographystyle{ieeetr}
\bibliography{references_cm}

@article{lasisi_fast_2022,
	title = {A {Fast} {Converging} {Resonance}-{Free} {Global} {Multi}-{Trace} {Method} for {Scattering} by {Partially} {Coated} {Composite} {Structures}},
	volume = {70},
	issn = {1558-2221},
	url = {https://ieeexplore.ieee.org/document/9818947},
	doi = {10.1109/TAP.2022.3187606},
	number = {10},
	urldate = {2024-12-09},
	journal = {IEEE Transactions on Antennas and Propagation},
	author = {Lasisi, Shakirudeen and Benson, Trevor M. and Gradoni, Gabriele and Greenaway, Mark and Cools, Kristof},
	month = oct,
	year = {2022},
	pages = {9534--9543},
}

@article{rao_electromagnetic_1982,
	title = {Electromagnetic scattering by surfaces of arbitrary shape},
	volume = {30},
	issn = {1558-2221},
	url = {https://ieeexplore.ieee.org/document/1142818},
	doi = {10.1109/TAP.1982.1142818},
	number = {3},
	urldate = {2024-12-09},
	journal = {IEEE Transactions on Antennas and Propagation},
	author = {Rao, S. and Wilton, D. and Glisson, A.},
	month = may,
	year = {1982},
	pages = {409--418},
}

@article{graglia_higher_1997,
	title = {Higher order interpolatory vector bases for computational electromagnetics},
	volume = {45},
	issn = {1558-2221},
	url = {https://ieeexplore.ieee.org/document/558649},
	doi = {10.1109/8.558649},
	number = {3},
	urldate = {2024-12-09},
	journal = {IEEE Transactions on Antennas and Propagation},
	author = {Graglia, R.D. and Wilton, D.R. and Peterson, A.F.},
	month = mar,
	year = {1997},
	pages = {329--342},
}

@article{claeys_electromagnetic_2012,
	title = {Electromagnetic scattering at composite objects : a novel multi-trace boundary integral formulation},
	volume = {46},
	issn = {2804-7214},
	shorttitle = {Electromagnetic scattering at composite objects},
	url = {http://www.numdam.org/item/M2AN_2012__46_6_1421_0/},
	doi = {10.1051/m2an/2012011},
	language = {en},
	number = {6},
	urldate = {2024-12-13},
	journal = {ESAIM: Mathematical Modelling and Numerical Analysis},
	author = {Claeys, Xavier and Hiptmair, Ralf},
	year = {2012},
	pages = {1421--1445},
}

@incollection{poggio_chapter_1973,
	series = {International {Series} of {Monographs} in {Electrical} {Engineering}},
	title = {{CHAPTER} 4 - {Integral} {Equation} {Solutions} of {Three}-dimensional {Scattering} {Problems}},
	isbn = {978-0-08-016888-3},
	url = {https://www.sciencedirect.com/science/article/pii/B9780080168883500088},
	urldate = {2024-12-13},
	booktitle = {Computer {Techniques} for {Electromagnetics}},
	publisher = {Pergamon},
	author = {Poggio, A. J. and Miller, E. K.},
	editor = {Mittra, R.},
	month = jan,
	year = {1973},
	doi = {10.1016/B978-0-08-016888-3.50008-8},
	pages = {159--264},
}

@article{buffa_dual_2007,
	title = {A dual finite element complex on the barycentric refinement},
	volume = {76},
	issn = {0025-5718, 1088-6842},
	url = {https://www.ams.org/mcom/2007-76-260/S0025-5718-07-01965-5/},
	doi = {10.1090/S0025-5718-07-01965-5},
	language = {English},
	number = {260},
	urldate = {2024-12-13},
	journal = {Math. Comp.},
	author = {Buffa, Annalisa and Christiansen, Snorre},
	month = oct,
	year = {2007},
	pages = {1743--1769},
}

@article{saad_gmres_1986,
	title = {{GMRES}: {A} {Generalized} {Minimal} {Residual} {Algorithm} for {Solving} {Nonsymmetric} {Linear} {Systems}},
	volume = {7},
	issn = {0196-5204},
	shorttitle = {{GMRES}},
	url = {https://epubs.siam.org/doi/10.1137/0907058},
	doi = {10.1137/0907058},
	number = {3},
	urldate = {2024-12-17},
	journal = {SIAM J. Sci. and Stat. Comput.},
	author = {Saad, Youcef and Schultz, Martin H.},
	month = jul,
	year = {1986},
	pages = {856--869},
}

@article{bourhis_high-order_2024,
	title = {High-{Order} {Quasi}-{Helmholtz} {Projectors}: {Definition}, {Analyses}, {Algorithms}},
	volume = {72},
	issn = {1558-2221},
	shorttitle = {High-{Order} {Quasi}-{Helmholtz} {Projectors}},
	url = {https://ieeexplore.ieee.org/document/10458016},
	doi = {10.1109/TAP.2024.3369752},
	number = {4},
	urldate = {2024-12-18},
	journal = {IEEE Transactions on Antennas and Propagation},
	author = {Bourhis, Johann and Merlini, Adrien and Andriulli, Francesco P.},
	month = apr,
	year = {2024},
	pages = {3572--3579},
}

@inproceedings{bourhis_novel_2024,
	title = {On a {Novel} {Calderón} {Preconditioning} {Strategy} {Based} on {High}-{Order} {Quasi}-{Helmholtz} {Projectors}},
	url = {https://ieeexplore.ieee.org/document/10701907},
	doi = {10.1109/ICEAA61917.2024.10701907},
	urldate = {2024-12-18},
	booktitle = {2024 {International} {Conference} on {Electromagnetics} in {Advanced} {Applications} ({ICEAA})},
	author = {Bourhis, Johann and Franzò, Damiano and Merlini, Adrien and Andriulli, Francesco P.},
	month = sep,
	year = {2024},
	pages = {700--703},
}

@article{andriulli_multiplicative_2008,
	title = {A {Multiplicative} {Calderon} {Preconditioner} for the {Electric} {Field} {Integral} {Equation}},
	volume = {56},
	issn = {1558-2221},
	url = {https://ieeexplore.ieee.org/document/4589072},
	doi = {10.1109/TAP.2008.926788},
	number = {8},
	urldate = {2024-12-19},
	journal = {IEEE Transactions on Antennas and Propagation},
	author = {Andriulli, Francesco P. and Cools, Kristof and Bagci, Hakan and Olyslager, Femke and Buffa, Annalisa and Christiansen, Snorre and Michielssen, Eric},
	month = aug,
	year = {2008},
	pages = {2398--2412},
}

@article{powell_optimal_2003,
	title = {Optimal {Preconditioning} for {Raviart}--{Thomas} {Mixed} {Formulation} of {Second}-{Order} {Elliptic} {Problems}},
	volume = {25},
	issn = {0895-4798},
	url = {https://epubs.siam.org/doi/abs/10.1137/S0895479802404428},
	doi = {10.1137/S0895479802404428},
	number = {3},
	urldate = {2025-03-10},
	journal = {SIAM J. Matrix Anal. Appl.},
	author = {Powell, Catherine Elizabeth and Silvester, David},
	month = jan,
	year = {2003},
    pages = {718--738},
}

@article{chu_surface_2003,
	title = {A surface integral equation formulation for low-frequency scattering from a composite object},
	volume = {51},
	issn = {1558-2221},
	url = {https://ieeexplore.ieee.org/document/1236103},
	doi = {10.1109/TAP.2003.817999},
	number = {10},
	urldate = {2025-04-04},
	journal = {IEEE Transactions on Antennas and Propagation},
	author = {Chu, Yunhui and Chew, Weng Cho and Zhao, Junsheng and Chen, Siyuan},
	month = oct,
	year = {2003},
	pages = {2837--2844},
}

@article{adrian_refinement-free_2019,
	title = {On a refinement-free {Calderón} multiplicative preconditioner for the electric field integral equation},
	volume = {376},
	issn = {0021-9991},
	url = {https://www.sciencedirect.com/science/article/pii/S0021999118306661},
	doi = {10.1016/j.jcp.2018.10.009},
	urldate = {2025-05-27},
	journal = {Journal of Computational Physics},
	author = {Adrian, S. B. and Andriulli, F. P. and Eibert, T. F.},
	month = jan,
	year = {2019},
	pages = {1232--1252},
}

@article{christiansen_preconditioner_2003,
	title = {A {Preconditioner} for the {Electric} {Field} {Integral} {Equation} {Based} on {Calderon} {Formulas}},
	volume = {40},
	issn = {0036-1429},
	url = {https://www.jstor.org/stable/4100917},
	number = {3},
	urldate = {2025-05-27},
	journal = {SIAM Journal on Numerical Analysis},
	author = {Christiansen, Snorre H. and Nédélec, Jean-Claude},
	year = {2003},
    pages = {1100--1135},
}

@article{valdes_high-order_2011,
	title = {High-order {Div}- and {Quasi} {Curl}-{Conforming} {Basis} {Functions} for {Calderón} {Multiplicative} {Preconditioning} of the {EFIE}},
	volume = {59},
	issn = {1558-2221},
	url = {https://ieeexplore.ieee.org/document/5705561},
	doi = {10.1109/TAP.2011.2109692},
	number = {4},
	urldate = {2025-06-04},
	journal = {IEEE Transactions on Antennas and Propagation},
	author = {Valdés, Felipe and Andriulli, Francesco P. and Cools, Kristof and Michielssen, Eric},
	month = apr,
	year = {2011},
	pages = {1321--1337},
}

@inproceedings{raviart_mixed_1977,
	address = {Berlin, Heidelberg},
	title = {A mixed finite element method for 2-nd order elliptic problems},
	isbn = {978-3-540-37158-8},
	doi = {10.1007/BFb0064470},
	language = {en},
	booktitle = {Mathematical {Aspects} of {Finite} {Element} {Methods}},
	publisher = {Springer},
	author = {Raviart, P. A. and Thomas, J. M.},
	editor = {Galligani, Ilio and Magenes, Enrico},
	year = {1977},
	pages = {292--315},
}

@article{masud_stabilized_2002,
	title = {A stabilized mixed finite element method for {Darcy} flow},
	volume = {191},
	issn = {0045-7825},
	url = {https://www.sciencedirect.com/science/article/pii/S0045782502003717},
	doi = {10.1016/S0045-7825(02)00371-7},
	number = {39},
	urldate = {2025-06-04},
	journal = {Computer Methods in Applied Mechanics and Engineering},
	author = {Masud, Arif and Hughes, Thomas J. R.},
	month = aug,
	year = {2002},
	pages = {4341--4370},
}

@article{napov_algebraic_2014,
	title = {Algebraic {Multigrid} for {Moderate} {Order} {Finite} {Elements}},
	volume = {36},
	issn = {1064-8275},
	url = {https://epubs.siam.org/doi/10.1137/130922616},
	doi = {10.1137/130922616},
	number = {4},
	urldate = {2025-06-05},
	journal = {SIAM J. Sci. Comput.},
	author = {Napov, Artem and Notay, Yvan},
	month = jan,
	year = {2014},
    pages = {A1678--A1707},
}

@phdthesis{merlini_unified_2019,
	type = {phdthesis},
	title = {Unified computational frameworks bridging low to high frequency simulations : fast and high fidelity modelling from brain to radio-frequency scenarios},
	shorttitle = {Unified computational frameworks bridging low to high frequency simulations},
	url = {https://theses.hal.science/tel-02466106},
	language = {en},
	urldate = {2025-06-05},
	school = {Ecole nationale supérieure Mines-Télécom Atlantique},
	author = {Merlini, Adrien},
	month = jan,
	year = {2019},
}

@article{andriulli_loop-star_2012,
	title = {Loop-{Star} and {Loop}-{Tree} {Decompositions}: {Analysis} and {Efficient} {Algorithms}},
	volume = {60},
	issn = {1558-2221},
	shorttitle = {Loop-{Star} and {Loop}-{Tree} {Decompositions}},
	url = {https://ieeexplore.ieee.org/document/6162946},
	doi = {10.1109/TAP.2012.2189723},
	number = {5},
	urldate = {2025-06-05},
	journal = {IEEE Transactions on Antennas and Propagation},
	author = {Andriulli, Francesco P.},
	month = may,
	year = {2012},
	pages = {2347--2356},
}

@article{livne_lean_2012,
	title = {Lean {Algebraic} {Multigrid} ({LAMG}): {Fast} {Graph} {Laplacian} {Linear} {Solver}},
	volume = {34},
	issn = {1064-8275},
	shorttitle = {Lean {Algebraic} {Multigrid} ({LAMG})},
	url = {https://epubs.siam.org/doi/10.1137/110843563},
	doi = {10.1137/110843563},
	number = {4},
	urldate = {2025-06-05},
	journal = {SIAM J. Sci. Comput.},
	author = {Livne, Oren E. and Brandt, Achi},
	month = jan,
	year = {2012},
    pages = {B499--B522},
}

@article{napov_algebraic_2012,
	title = {An {Algebraic} {Multigrid} {Method} with {Guaranteed} {Convergence} {Rate}},
	volume = {34},
	issn = {1064-8275},
	url = {https://epubs.siam.org/doi/10.1137/100818509},
	doi = {10.1137/100818509},
	number = {2},
	urldate = {2025-06-05},
	journal = {SIAM J. Sci. Comput.},
	author = {Napov, Artem and Notay, Yvan},
	month = jan,
	year = {2012},
    pages = {A1079--A1109},
}

@article{chen_analysis_2001,
	title = {Analysis of low frequency scattering from penetrable scatterers},
	volume = {39},
	issn = {1558-0644},
	url = {https://ieeexplore.ieee.org/document/917883},
	doi = {10.1109/36.917883},
	number = {4},
	urldate = {2025-10-31},
	journal = {IEEE Transactions on Geoscience and Remote Sensing},
	author = {Chen, S.Y. and Chew, Weng Cho and Song, J.M. and Zhao, Jun-Sheng},
	month = apr,
	year = {2001},
	pages = {726--735},
}

@article{zhao_integral_2000,
	title = {Integral equation solution of {Maxwell}'s equations from zero frequency to microwave frequencies},
	volume = {48},
	issn = {1558-2221},
	url = {https://ieeexplore.ieee.org/document/899680},
	doi = {10.1109/8.899680},
	number = {10},
	urldate = {2025-10-31},
	journal = {IEEE Transactions on Antennas and Propagation},
	author = {Zhao, Jun-Sheng and Chew, Weng Cho},
	month = oct,
	year = {2000},
	pages = {1635--1645},
}

@inproceedings{bogaert_low_2011,
	title = {Low frequency scaling of the mixed {MFIE} for scatterers with a non-simply connected surface},
	url = {https://ieeexplore.ieee.org/document/6046467},
	doi = {10.1109/ICEAA.2011.6046467},
	urldate = {2025-10-31},
	booktitle = {2011 {International} {Conference} on {Electromagnetics} in {Advanced} {Applications}},
	author = {Bogaert, I. and Cools, K. and Andriulli, F. P. and De Zutter, D.},
	month = sep,
	year = {2011},
	pages = {951--954},
}

@article{bogaert_low-frequency_2014,
	title = {Low-{Frequency} {Scaling} of the {Standard} and {Mixed} {Magnetic} {Field} and {Müller} {Integral} {Equations}},
	volume = {62},
	issn = {1558-2221},
	url = {https://ieeexplore.ieee.org/document/6678539},
	doi = {10.1109/TAP.2013.2293783},
	number = {2},
	urldate = {2025-10-31},
	journal = {IEEE Transactions on Antennas and Propagation},
	author = {Bogaert, Ignace and Cools, Kristof and Andriulli, Francesco P. and Bağcı, Hakan},
	month = feb,
	year = {2014},
	pages = {822--831},
}

@article{merlini_magnetic_2020,
	title = {Magnetic and {Combined} {Field} {Integral} {Equations} {Based} on the {Quasi}-{Helmholtz} {Projectors}},
	volume = {68},
	issn = {1558-2221},
	url = {https://ieeexplore.ieee.org/document/8963862},
	doi = {10.1109/TAP.2020.2964941},
	number = {5},
	urldate = {2025-10-31},
	journal = {IEEE Transactions on Antennas and Propagation},
	author = {Merlini, Adrien and Beghein, Yves and Cools, Kristof and Michielssen, Eric and Andriulli, Francesco P.},
	month = may,
	year = {2020},
	pages = {3834--3846},
}

@article{beghein_low-frequency_2017,
	title = {On a {Low}-{Frequency} and {Refinement} {Stable} {PMCHWT} {Integral} {Equation} {Leveraging} the {Quasi}-{Helmholtz} {Projectors}},
	volume = {65},
	issn = {1558-2221},
	url = {https://ieeexplore.ieee.org/document/8007283},
	doi = {10.1109/TAP.2017.2738061},
	number = {10},
	urldate = {2025-10-31},
	journal = {IEEE Transactions on Antennas and Propagation},
	author = {Beghein, Yves and Mitharwal, Rajendra and Cools, Kristof and Andriulli, Francesco P.},
	month = oct,
	year = {2017},
	pages = {5365--5375},
}

\end{document}